\documentclass[fleqn,usenatbib]{mnras}

\usepackage{newtxmath}

\usepackage[T1]{fontenc}
\usepackage{ae,aecompl}

\usepackage{graphicx}	
\usepackage{amsmath}	
\usepackage{amssymb}	

\title[RAVE-Gaia streaming motions]{Is the Milky Way still breathing? RAVE-Gaia streaming motions}

\author[I. Carrillo et al.]{
I. Carrillo,$^{1}$\thanks{E-mail: icarrillo@aip.de}
I. Minchev,$^{1}$
G. Kordopatis,$^{1, 2}$
M. Steinmetz,$^{1}$
J. Binney,$^{3}$
\newauthor
F. Anders,$^{1}$
O. Bienaym\'{e},$^{4}$
J. Bland-Hawthorn,$^{5}$
B. Famaey,$^{4}$
K.C. Freeman,$^{6}$
\newauthor
G. Gilmore,$^{7}$
B. K. Gibson,$^{8}$
E.~K.~Grebel,$^{9}$
A. Helmi,$^{10}$
A. Just,$^{9}$
A. Kunder,$^{1, 11}$
\newauthor
P. McMillan,$^{12}$
G. Monari,$^{1, 13}$
U. Munari,$^{14}$
J. Navarro,$^{15}$
Q. A. Parker,$^{16, 17}$
\newauthor
W. Reid,$^{18, 19}$
G. Seabroke,$^{20}$
S. Sharma,$^{5}$
A. Siebert,$^{4}$
F. Watson,$^{21}$
\newauthor
J. Wojno,$^{1}$
R. F. G. Wyse,$^{22}$
T. Zwitter$^{23}$
\\
$^{1}$Leibniz Institut f\"{u}r Astrophysik Potsdam (AIP), An der Sterwarte 16, D-14482 Potsdam, Germany\\
$^{2}$Universit\'{e} C\^{o}te d'Azur, Observatoire de la C\^{o}te d'Azur, CNRS, Laboratoire Lagrange,\\ Bd de l'Observatoire, CS 34229, 06304, Nice cedex 4, France\\
$^{3}$Rudolf Peierls Centre for Theoretical Physics, Keble Road, Oxford, OX1 3NP, UK\\
$^{4}$Observatoire astronomique de Strasbourg, Universit\'{e} de Strasbourg, CNRS, UMR 7550, 11 rue de  l'Universit\'{e}, F-67000 Strasbourg, France\\
$^{5}$Sydney Institute for Astronomy, School of Physics A28, University of Sydney, NSW 2006, Australia\\
$^{6}$Research School of Astronomy \& Astrophysics, The Australian National University\\
$^{7}$Institute of Astronomy, University of Cambridge, Madingley Road, Cambridge, CB3 0HA, UK\\
$^{8}$E.A. Milne Centre for Astrophysics, University of Hull, Hull, HU6 7RX, UK\\
$^{9}$Astronomisches Rechen-Institut, Zentrum f\"{u}r Astronomie der Universit\"{a}t Heidelberg, M\"{o}nchhofstr. 12--14, 69120 Heidelberg, Germany\\
$^{10}$Kapteyn Astronomical Institute, University of Groningen, PO Box 800, NL-9700 AV Groningen, the Netherlands\\
$^{11}$Saint Martin's University, 5000 Abbey Way SE, Lacey, WA 98503, USA\\
$^{12}$Lund Observatory, Lund University, Department of Astronomy and Theoretical Physics, Box 43, SE-22100, Lund, Sweden\\
$^{13}$The Oskar Klein Centre for Cosmoparticle Physics, Department of Physics, Stockholm University, AlbaNova, 10691 Stockholm, Sweden\\
$^{14}$INAF National Institute of Astrophysics, Astronomical Observatory of Padova, 36012 Asiago, Italy\\
$^{15}$Senior CIfAR Fellow. Department of Physics and Astronomy. University of Victoria. Victoria, BC Canada V8P5C2\\
$^{16}$Department of Physics, the University of Hong Kong, Hong Kong SAR, China\\
$^{17}$The Laboratory for Space Research, the university of Hong kong, Hong Kong SAR, China\\
$^{18}$Department of Physics and Astronomy, Macquarie University, Sydney, NSW 2109, Australia\\
$^{19}$Western Sydney University, Locked Bag 1797, Penrith South, NSW 2751, Australia\\
$^{20}$Mullard Space Science Laboratory, University College London, Holmbury St Mary, Dorking, RH5 6NT, UK\\
$^{21}$Australian Astronomical Observatory PO Box 915, North Ryde, NSW 1670, Australia\\
$^{22}$Department of Physics and Astronomy, Johns Hopkins University, 3400 N. Charles St, Baltimore, MD 21218, USA\\
$^{23}$Faculty of Mathematics and Physics, University of Ljubljana, Jadranska 19, 1000 Ljubljana, Slovenia}

\date{Accepted XXX. Received YYY; in original form ZZZ}

\pubyear{2017}

\begin{document}
\label{firstpage}
\pagerange{\pageref{firstpage}--\pageref{lastpage}}
\maketitle

\begin{abstract}
We use data from the Radial Velocity Experiment (RAVE)
and the Tycho-Gaia astrometric solution catalogue (TGAS) to compute 
the velocity fields yielded by the radial ($V_R$), azimuthal ($V_{\phi}$)
and vertical ($V_z$) components of associated 
Galactocentric velocity. We search in particular for variation in all three velocity
components with distance above and below the disc midplane, as well as how
each component of $V_z$ (line-of-sight and tangential velocity projections)
modifies the obtained vertical structure. To study the dependence of velocity
on proper motion and distance we use two main samples: a RAVE sample
including proper motions from the Tycho\,-2, PPMXL and UCAC4 catalogues, and
a RAVE-TGAS sample with inferred distances and proper motions from the TGAS
and UCAC5 catalogues. In both samples, we identify asymmetries in
$V_R$ and $V_z$. Below the plane we find the largest radial gradient to be
$\partial V_R /\partial R = -7.01\pm 0.61$\,km\,s$^{-1}$\,kpc$^{-1}$, in
agreement with recent studies. Above the plane we find a similar gradient
with $\partial V_R /\partial R= -9.42 \pm 1.77$\,km\,s$^{-1}$\,kpc$^{-1}$. By
comparing our results with previous studies, we find that the structure in
$V_z$ is strongly dependent on the adopted proper motions. Using the Galaxia
Milky Way model, we demonstrate that distance uncertainties can create
artificial wave-like patterns. In contrast to previous suggestions of a
breathing mode seen in RAVE data, our results support a combination of
bending and breathing modes, likely generated by a combination of external or
internal and external mechanisms.
\end{abstract}


\defcitealias{wobbly}{W13}
\section{Introduction}
To a first approximation, the Milky Way disc is assumed to be axisymmetric
and in equilibrium (e.g. \citealp{nonax1}; \citealp{nonax2};
\citealp{nonax3}). However, in the last two decades with the acquisition of high-quality
spectroscopic and astrometric data, and expansion of the surveyed volume
around the Sun, non-negligible deviations from axisymmetry have become
apparent. One indication of such asymmetries came
from the clear identification in data from the Hipparcos astrometric
satellite \citep{hipparcos} of overdensities in the velocity space of local
stars (\citealp{1998Dehnen}; \citealp{Chereul}; \citealp{Asiain}). Subsequent
analysis of the local velocity field have revealed that the most prominent
moving groups (or streams) in the solar neighbourhood probably have a
dynamical origin since within a given stream there is a mixture of chemical abundances and
ages (e.g. \citealp{Famaey2005}; \citealp{2008Antoja};
\citealp{Famaey2008}). \citet{2000Dehnen} and \citet{Fuxbar} showed that the
Galactic bar can successfully reproduce the Hercules stream if the Sun is
situated close to the bar's outer Lindblad resonance (OLR), while a solar
position near the inner ultra-harmonic 4:1 resonance of a two-armed spiral
density wave can similarly create resonant structures consistent with the
Pleiades/Hyades and Coma Berenices moving groups (\citealp{Quillenspiral};
\citealp{Pompeia2011}). While internal perturbations can explain the
low-velocity moving groups, high-velocity streams (such as Arcturus, which
lags the local standard of rest by $\sim100$\,km\,s$^{-1}$;
\citealp{Williams2009}), have been related to external perturbations from an
infalling satellite galaxy (\citealp{Minchev2009}; \citealp{Gomez2012a},
\citeyear{Gomez2012b}; \citealp{Elena2016}). A crucial diagnostic in understanding what causes
these streaming motions is how structure of the velocity space varies with
location in the disc (\citealp{Dehnen2001}; \citealp{Minchev2010};
\citealp{Antoja2014}; \citealp{McMillan2013}; \citealp{Monari2017}).

Another manifestation of non-axisymmetries in the Milky Way disc is in-plane
stellar streaming motions detected in the extended solar neighborhood. Using
line-of-sight velocities measured in the Radial Velocity Experiment (RAVE;
\citealp{DR1}) \citet{siebertgradient} measured a gradient in the mean
Galactocentric radial velocity ($\left |\partial \left \langle V_R \right
\rangle/\partial R \right |\gtrsim 3$\,km\,s$^{-1}$\,kpc$^{-1}$).
\citet{Siebert2012} attributed this gradient to a two-armed spiral
perturbation in which the Sun is again close to the inner ultra-harmonic 4:1
resonance. More recently, \citet{Monari2014} used test-particle simulations
to show that the radial gradient found in RAVE could alternatively be caused
by the Galactic bar.

In addition to stellar bulk motions in the disc plane, structure has been
found also in the direction perpendicular to the Galactic disc.
\citet{Widrow2012} and \citet[][hereafter W13]{wobbly} used SEGUE
\citep{Yanny2009} and RAVE, respectively, to detect a wave-like pattern in
the mean vertical velocity of stars near the Sun. Similar vertical
asymmetries were also found by \citet{Carlin2013} using data from the LAMOST
survey \citep{LAMOST} and \citet{Xu2015} with data from the Sloan Digital Sky
Survey \citep{SDSSdr8}. 

While the radial velocity gradient has been associated with internal
perturbations, the origin of vertical velocity structure is debatable.
\citet{Gomez2013} attributed the vertical patterns to the passing of the
Sagittarius dwarf galaxy \citep{Ibata1994}, while \citet{Widrow2014} noted
that they could also be caused by a dark matter subhalo. In addition to external
perturbations, vertical streaming motions have now been shown to result also
from internal mechanisms such as the Galactic bar \citep{Monari2015} and
spiral arms (\citealp{2014Faure}; \citealp{2016Monari}). However, according
to \citet{Monari2015}, the Galactic bar is unlikely to induce mean vertical
motions greater than $\sim 0.5$\,km\,s$^{-1}$ in the outer disc, and
therefore does not explain the observed vertical motions in the solar
neighbourhood. 

Another possible explanation of the observed vertical motions is the Galactic warp. The Galactic disc becomes warped beyond the solar circle: the H{\footnotesize I} disc at $R\ga10\,$kpc is warped such that the Sun lies near the line of nodes and the $z$ coordinate of the centre of the gas layer increases in the direction of
Galactic rotation (\citealt{BinneyM}; \citealp{Levine2006};
\citealp{Kalberla2007}). Consequently, if (as is likely) the pattern speed
of the warp is smaller than the circular frequency of the Sun, gas at $R>R_0$
and the Sun's azimuth should be moving downwards ($V_z<0$). A warp in the
stellar disc has been detected from $R_0$ outwards (e.g.\
\citealp{1998Dehnen}; \citealp{Drimmel2001}; \citealp{Robin2008}).
Nonetheless, \citet{Poggio2017} used OB stars and proper motions from
Hipparcos and {\it Gaia} DR1 \citep{GaiaDR1first} to develop a simple model
of a stable long-lived warp and found that vertical motions in the disc
cannot be explained by the warp, suggesting that the warp's kinematic signal
could be overwhelmed by other systematic motions.

More generally, the analysis of 16 high-resolution, fully cosmological simulations of the evolution of
individual Milky Way-sized galaxies led \citet{Gomez2016} to conclude that
vertical disc asymmetries are ubiquitous in $\Lambda$CDM cosmology. In fact,
$\sim70\%$ of the analysed simulations exhibited strong vertical patterns,
with amplitudes exceeding 2\,kpc from the disc midplane. Hence
vertical motions are predicted by $\Lambda$CDM cosmology.

Using RAVE red-clump stars and a compilation of proper motions,
\citetalias{wobbly} studied the 3D velocity distribution in the extended
solar neighbourhood, focusing on north-south differences, and detected a
rarefaction-compression behaviour in the vertical velocity pattern. Such
a pattern, which has odd parity in the $V_z$ distribution and even
parity in the density distribution, is known as a breathing mode. In
contrast, even parity in $V_z$ with odd parity in the density
distribution is known as a bending mode. By identifying the observed mode, we
hope to constrain the nature of the exciting perturber. Breathing modes have
been found to be induced naturally by the Galactic bar and spiral arms
(\citealp{2014Faure}; \citealp{Monari2015}; \citealp{2016Monari}), whereas
bending modes are attributed mainly to external perturbations. While more
recently, \citet{Chequers2017} found that bending waves could also
arise without excitation by a satellite/merging event, \citet{Widrow2014}
used a toy-model simulation of disc-satellite interactions to show that a
passing satellite galaxy could produce bending or breathing modes depending
on the vertical velocity of the satellite as it passes through the Galactic
disc. This shows that the picture can actually be rather complex, as external
perturbations also generate internal spiral perturbations which can in turn
excite breathing modes. While both bending and breathing modes
can arise from external interactions, the observations so far have been found
consistent with a breathing mode. Another interpretation was found by
\citet{delaVega2015}, who showed that phase-wrapping in the disc following
the passage of a satellite, can look very similar to a bending or a breathing
mode.

ESA's mission {\it Gaia} \citep{Gaia_mission} is acquiring highly accurate
parallaxes and proper motions for over a billion sources brighter than
magnitude 20.7 in the G band. In this paper, we use early mission data from
the first data release ({\it Gaia} DR1) combined with the line-of-sight
velocities of the fifth data release from RAVE (DR5; \citealp{RAVEDR5}) to
study the three-dimensional velocity distribution of stars in the extended
solar neighbourhood. We extend the analysis of \citetalias{wobbly} by
increasing the number of stars through inclusion of stars that are not in the
red-clump and using better proper motions and distances. Since
uncertainties in velocity fields are dominated by errors in distance and
proper motion (RAVE uncertainty in line-of-sight velocity
$<2$\,km\,s$^{-1}$), our goal is to use proper motions from {\it Gaia} to
scrutinize with higher accuracy the vertical velocity pattern of the extended
solar neighbourhood. In this way we aim to determine whether the Milky Way
exhibits a breathing or a bending mode.

This paper is structured as follows: In Section~\ref{sec:data_coord}, we
introduce our coordinate conventions and explain how we selected stars. In
Section~\ref{sec:DR5_wobbly}, we present the radial, azimuthal and vertical
velocity distribution obtained by \citetalias{wobbly} and the velocity
distributions obtained from RAVE DR5 and different proper-motion catalogues.
In Section~\ref{sec:Vz_pattern} we study the composition of the vertical
velocity pattern and how errors affect its structure. In
Section~\ref{sec:RAVE-TGAS}, we examine the differences between RAVE DR5 and
samples that use data from DR5 combined with the Tycho-Gaia astrometric
solution catalogue (RAVE-TGAS samples) and present the north-south asymmetries
obtained using the most accurate estimates of distance and proper motion.
Finally, Section~\ref{sec:Concl} contains a summary of our results and
conclusions.

\section{Coordinate systems \& Data selection}
\label{sec:data_coord}

\subsection{Coordinate systems} \label{sec:cylindrical} 

We compute the heliocentric rectangular components of the Galactic space
velocity $U$, $V$ and $W$ using the right-handed coordinate system, with $U$
positive towards the Galactic centre, $V$ positive in the direction of
Galactic rotation and $W$ positive towards the north Galactic pole. The
method used to derive the Galactic space velocities is described in detail in
\citet{1987}. The Galactocentric cylindrical velocity components ($V_R$,
$V_{\phi}$ and $V_z$) are computed following the coordinate transformation
given in Appendix A of \citetalias{wobbly}.

For the solar Galactocentric distance,
we adopt $R_0=8$\,kpc \citep{DistanceGC} and use the estimate of
the peculiar velocity of the Sun obtained by \citet{binney}:
\begin{equation}
\label{eq:Sunpeculiar}
(U,V,W)_{\odot}=(11.1,\,12.24,\,7.25)\,\text{km s}^{-1}
\end{equation}
With these values and the proper motion of Sagittarius\,A*, $\mu_{l_\mathrm{
Sgr\,A^*}}\,=6.379$ mas\, yr$^{-1}$ \citep{Reid2004}, we obtain:
\begin{equation}
V_{\odot}+V_\mathrm{LSR} = 4.74\,R_0\,\mu_{l_\mathrm{ Sgr\,A^*}},
\end{equation} 
which yields a value of $V_\mathrm{LSR}\sim230$\,km\,s$^{-1}$ for the
circular velocity of the local standard of rest (LSR). We also tested our
results with $R_0=8.3$\,kpc and $V_\mathrm{LSR}=240$\,km\,s$^{-1}$. The changes in $V_R$ and $V_{\phi}$ were negligible, and since $V_z$ is independent of $V_\mathrm{LSR}$ it suffers no changes aside from the position in $R$.

\begin{figure*}
	\includegraphics[scale=.93]{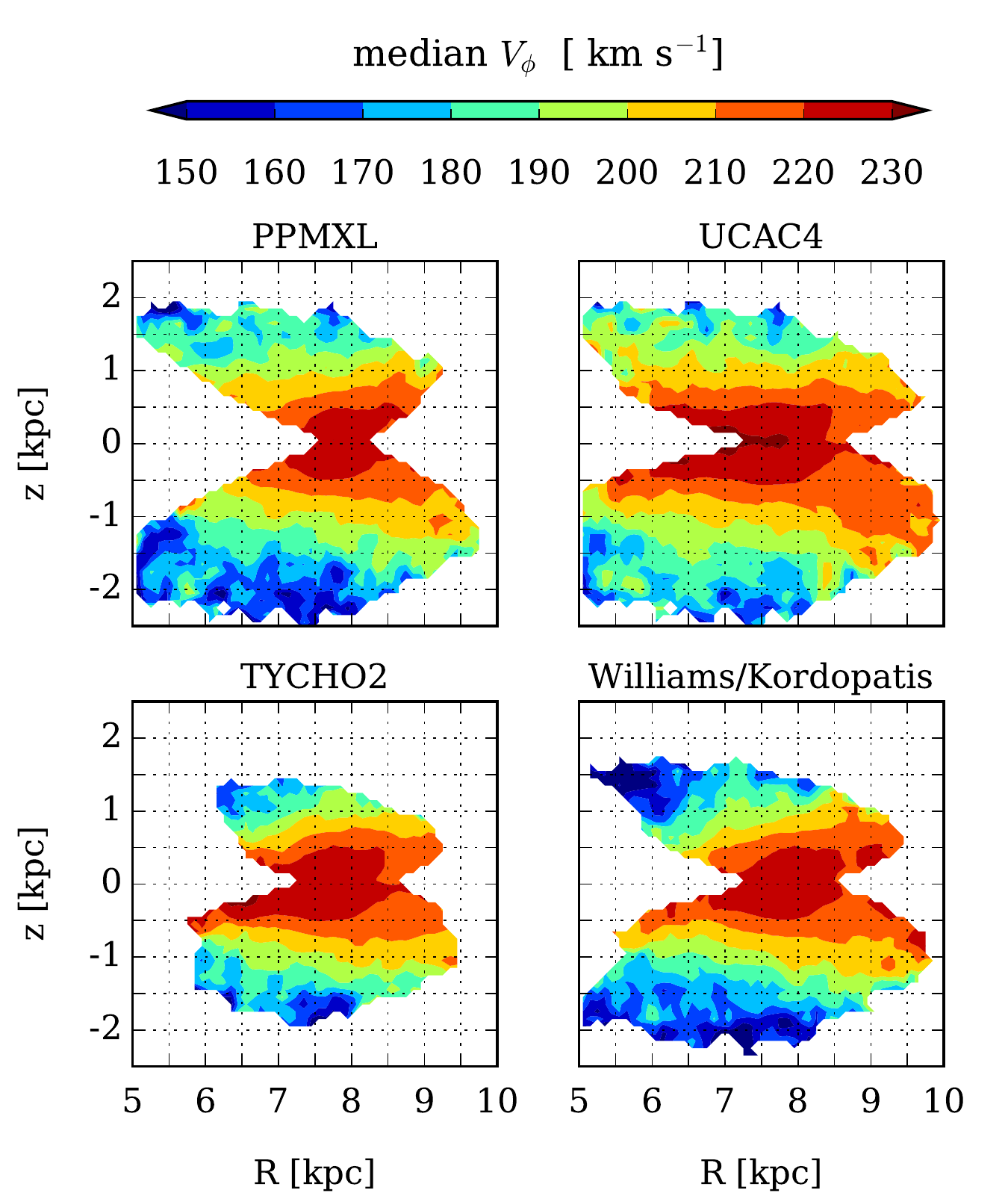}
	\caption{Median azimuthal velocity fields obtained using different
proper-motion catalogues, as indicated on top of each panel. Each velocity
field is shown in (0.1\,kpc)$^2$ pixels covering a box of up to (5 kpc)$^2$
around the Sun, which we locate at ($R,z$)=(8,0)\,kpc. All pixels are
smoothed by computing the median of the velocities of all stars located in a
square of size (0.2\,kpc)$^2$ with a minimum of 50 stars. For comparison
purposes, the bottom right panel shows the velocity field obtained by
\citet{GK2013} which is very similar to the one obtained using only RC stars by \citet{wobbly}.
The velocity fields of all proper motions are consistent with a decrease of
$V_{\phi}$ with increasing distance from the plane.} 
	\label{fig:DR5_vphi}
\end{figure*}
\begin{figure*}
	\includegraphics[scale=.93]{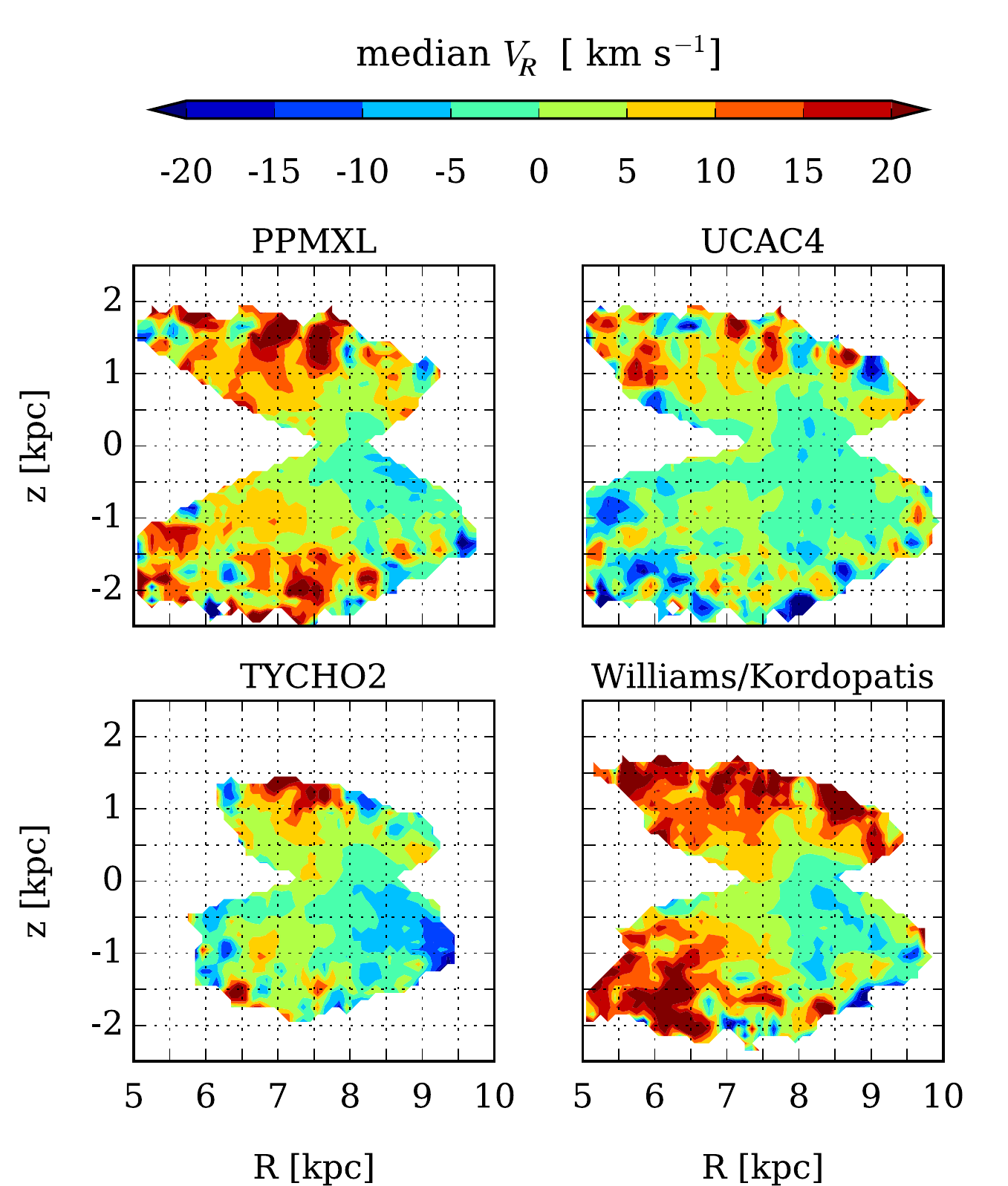}\\
	\caption{Same as Fig.\,\ref{fig:DR5_vphi} but for $V_R$. Independent
of the proper motion catalogue used, the velocity field above the plane
is structured similarly, with mostly positive velocities everywhere. In contrast,
below the plane the structure varies significantly between proper-motions catalogues.}
	\label{fig:DR5_vr}
\end{figure*}
\begin{figure}
	\includegraphics[scale=.45]{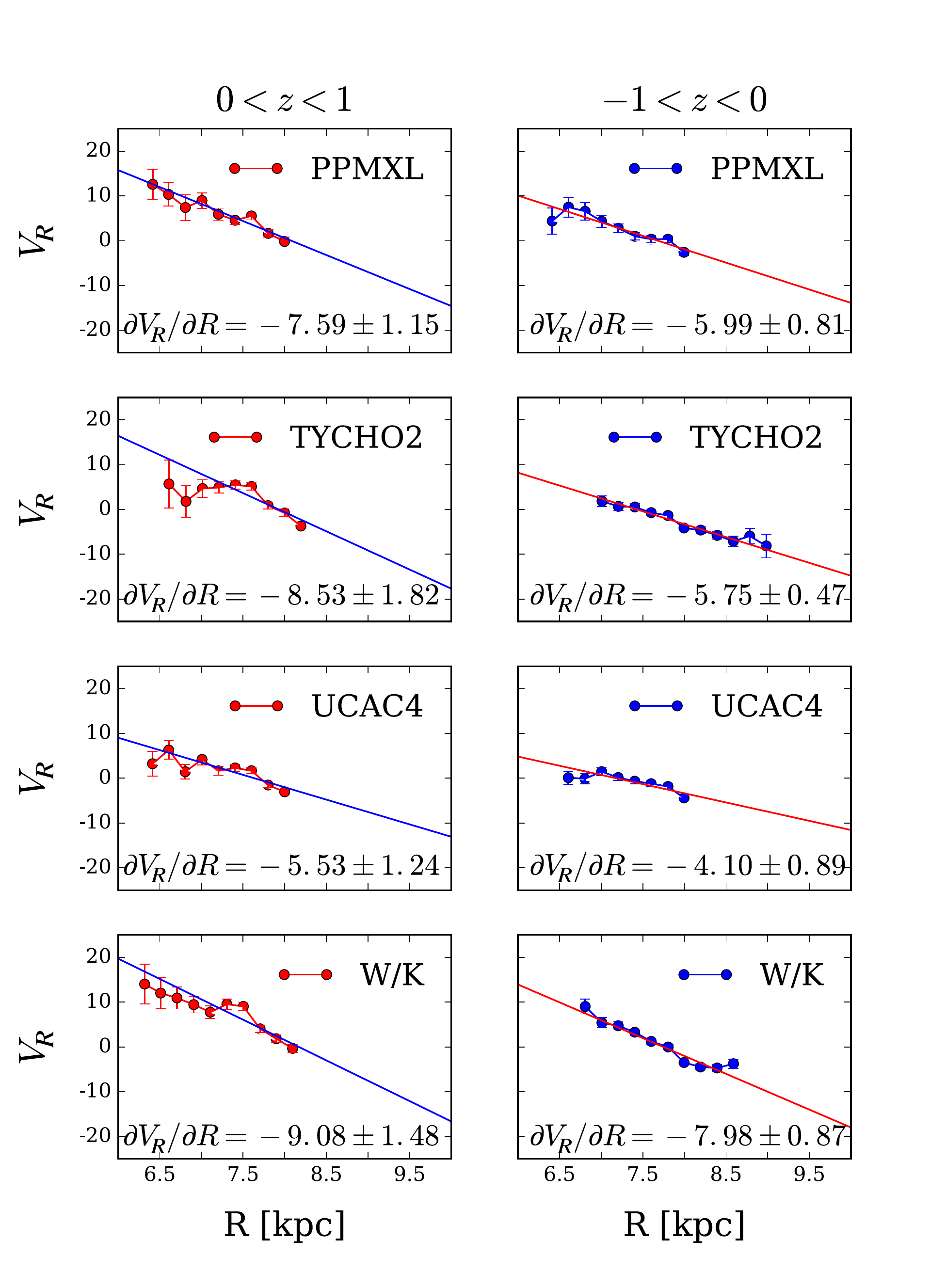}
	\caption{Largest negative gradient in $V_R$ estimated in each
proper motion catalogue for stars above (left) and below the plane (right)
computed in bins of 0.2\,kpc using the shown ($R,z$) planes.}
	\label{fig:Gradients}
\end{figure}
\begin{figure*}
	\includegraphics[scale=.93]{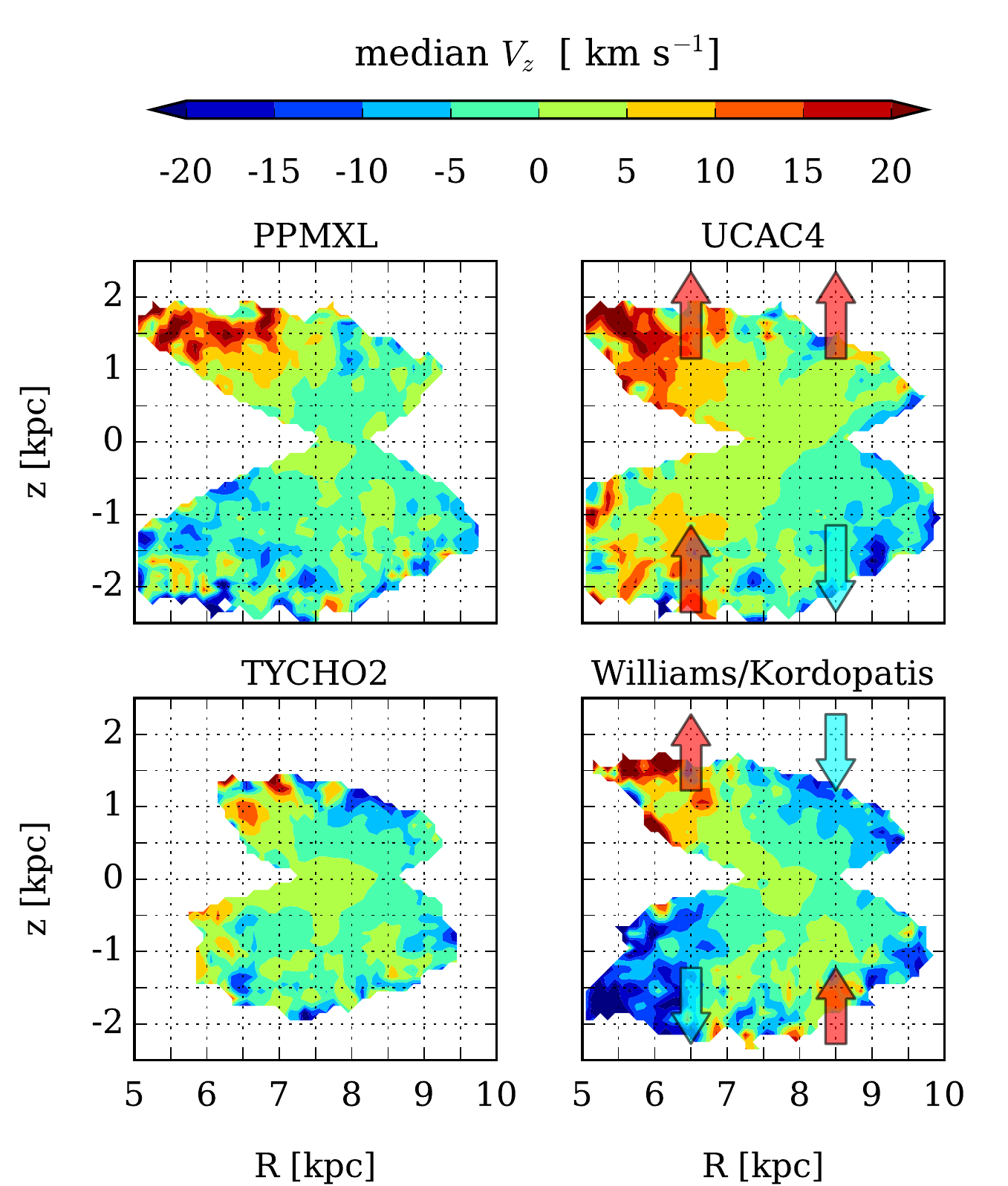}\\
	\caption{Same as Fig.\,\ref{fig:DR5_vphi} but for $V_z$. The trend in
	vertical velocity varies significantly with proper-motion catalogue,
especially below the plane. The arrows in the bottom right panel show the
direction of vertical motion consistent with a breathing mode.
While some catalogues agree with the breathing mode observed by
Williams/Kordopatis, others display a bending mode or a combination of
bending and breathing modes not previously seen (top right panel).}
	\label{fig:DR5_vz}
\end{figure*}
\begin{figure*}
	\includegraphics[width=\textwidth]{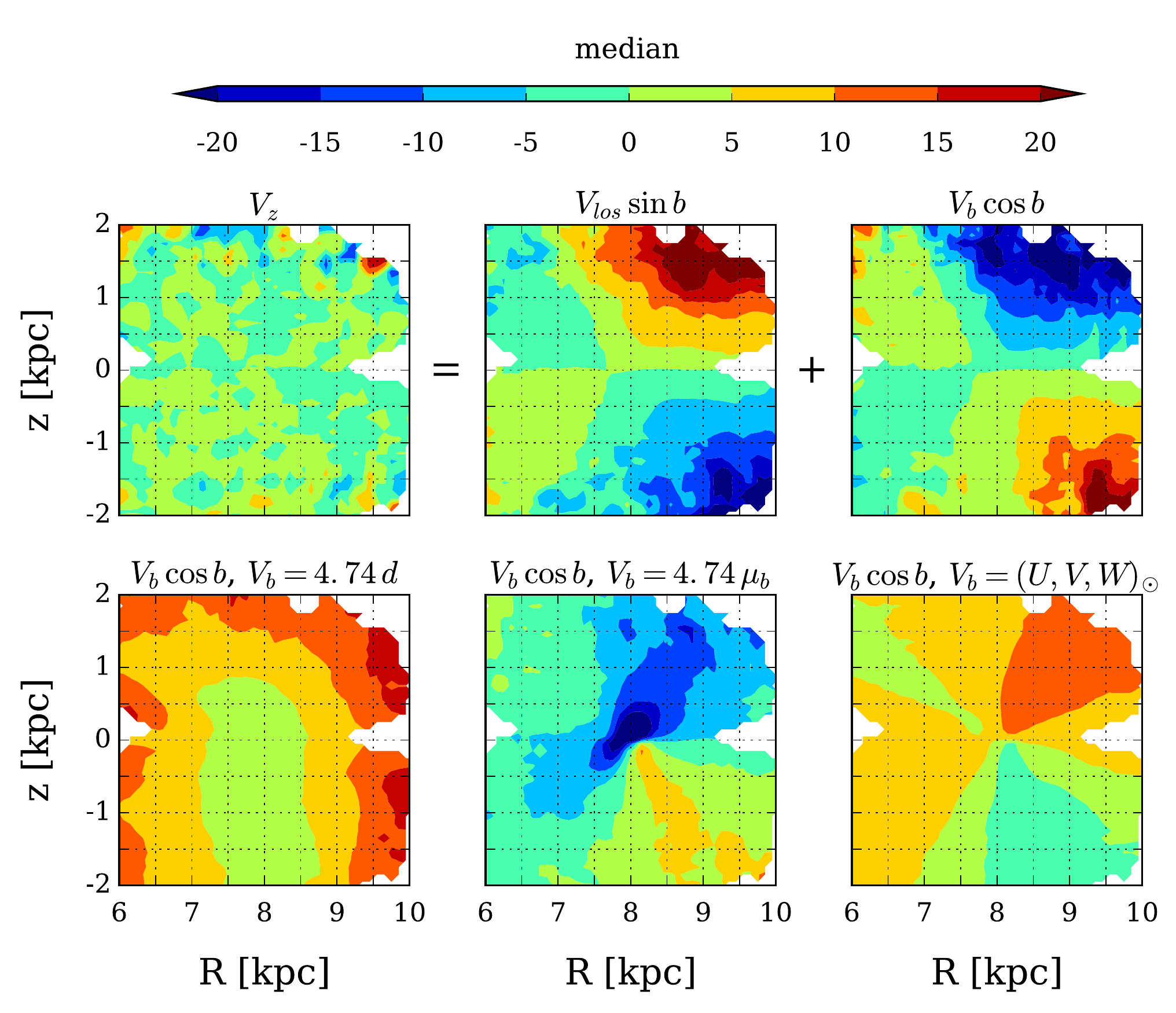}
	\caption{Decomposition of the median value of $V_z$ obtained from the
RAVE-like sample from Galaxia assuming a magnitude limit $0 < I < 13$, with
no colour restriction, where RAVE-like errors were applied to stellar
parameters. The velocity fields are shown in (0.1\,kpc)$^2$ pixels covering a
box of up to (4 kpc)$^2$ around the Sun. The top panels display $V_z$ and its
two main components based on Eq.\,\ref{eq:vz}. The bottom panels show the
contributions to $V_b\cos b$ (distance, proper motion and solar motion) from
Eq.\,\ref{eq:vb}. These panels indicate how each component could affect the map of $V_z$.
The transverse velocity depends mainly on the proper motion with an amplitude that increases
with distance.}
	\label{fig:Galaxia}
\end{figure*}
\begin{figure*}
	\includegraphics[width=\textwidth]{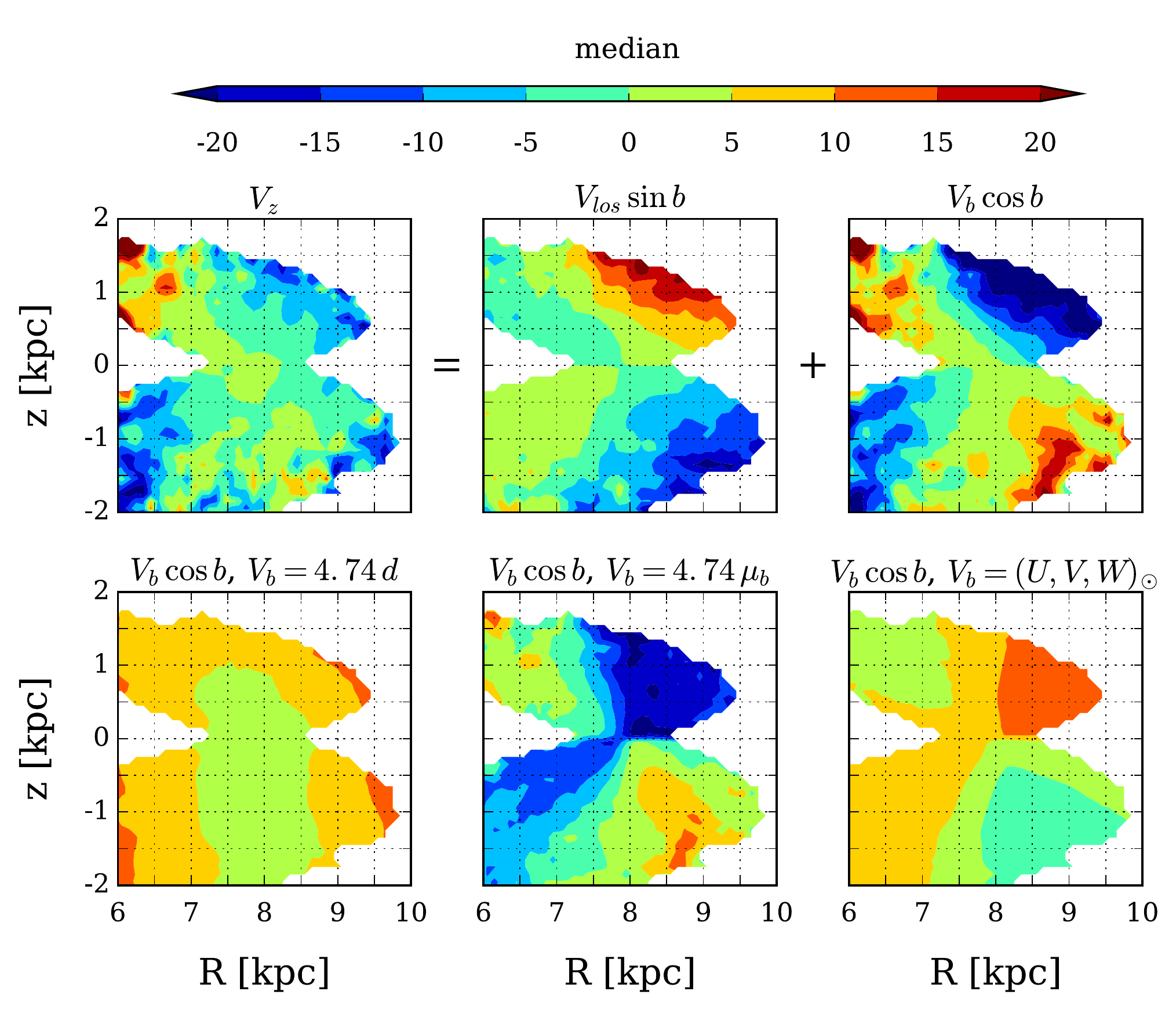}
	\caption{Same as Fig.\,\ref{fig:Galaxia} but for the map of $V_z$
obtained by Williams/Kordopatis.}
	\label{fig:GK_comp}
\end{figure*}
\begin{figure*}
	\includegraphics[width=\textwidth]{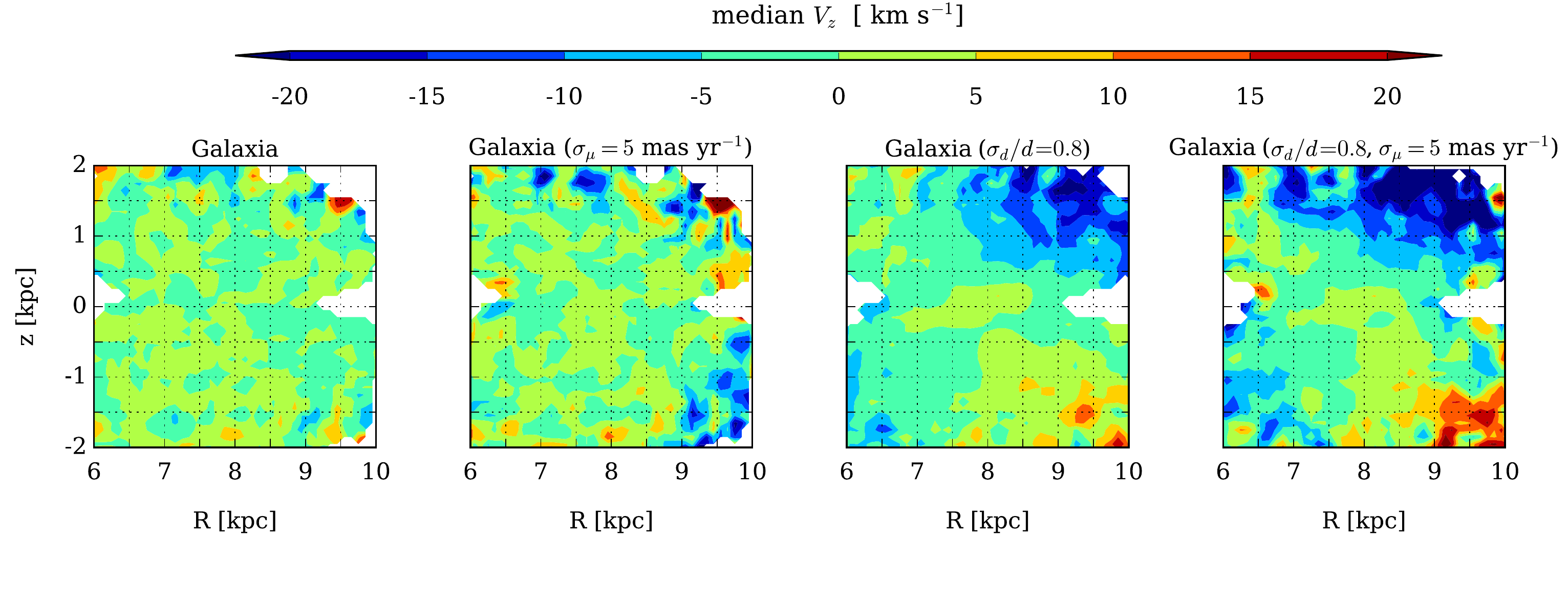}
	\caption{Median $V_z$ obtained from the Galaxia RAVE-like sample together with the effects of adding random errors in $d$ and $\mu$ as described
in the text. The second panel shows that by adding a random error in $\mu$
the $V_z$ pattern displays some clumps, while the third panel shows that
adding distance errors substantially changes the map. The combination of both
errors in the last panel results in a pattern similar to the one obtained by
Williams/Kordopatis.}
	\label{fig:Galaxia_errors}
\end{figure*}

\subsection{RAVE data selection} \label{sec:RAVE_data}

RAVE is a spectroscopic magnitude-limited ($9 \lesssim I \lesssim 12$)
southern hemisphere survey, which collected data from 2003 to 2013. It
acquired line-of-sight velocities and stellar parameters (effective
temperature, surface gravity, overall metallicity) in order to probe the
chemical and dynamical evolution of the Milky Way. In this paper we use its
latest data release, Data Release 5 \citep{RAVEDR5}. DR5 contains
520,781 spectra of 457,588 unique stars. It includes observations
that were previously discarded, resulting in $ \sim 30,000$ more spectra than
the fourth data release (DR4; \citealp{DR4}). DR5 also improves on the
distance pipeline of DR4 \citep{binneydistances}, especially as regards
metal-poor stars, and applies a new calibration of its stellar parameters
that improves their accuracy by up to $15\%$ compared to DR4 (See Section~6
in \citealp{RAVEDR5} for details). The proper motions in DR5 are taken from
the Tycho\,-2 \citep{Tycho2}, PPMXL \citep{PPMXL} and UCAC4 \citep{UCAC4}
catalogues.

For our data selection (RAVE sample), we exclude stars without measured proper
motions and stars with a proper motion error
$\sigma_{\mu_{\alpha}}, \sigma_{\mu_{\delta}} >15$ mas\, yr$^{-1}$ (catalogue
independent) to remove outliers. Further cuts applied were to include only
stars with a signal-to-noise ratio $\mathrm{SNR_K} >20$ and to remove stars
whose spectral morphological flag indicates peculiar spectra (c1,c2,c3 $\neq$
'n'); these cuts improve the quality of our data and remove most
spectroscopic binaries, spectra with continuum abnormalities and other
unusual spectra. We also removed stars with an error of heliocentric radial
velocity $\mathrm{eHRV}>10$\,km\,s$^{-1}$, cool dwarfs (log\,$g>3.5$ and
$T_{\mathrm{eff}}<4000$\,K), metal poor stars ($\mathrm{[M/H]}<-1.2$\,dex),
hot stars ($T_{\mathrm{eff}}>7400$\,K) and spectra for which atmospheric
parameters could not be determined reliably ($\mathrm{Algoconv}=1$), since in that
case, the derived distances (obtained through isochrone fitting, see
\citealp{binneydistances}) were not reliable either.

Finally, stars at Galactic latitude $ \left | b \right |< 10^{\circ} $ were
removed due to uncertain interstellar extinctions possibly affecting the
derived distances. Our data selection applies to all proper-motion
catalogues, with the exception of the Galactic latitude cut for the PPMXL
catalogue, from which we remove all stars with $ \left | b \right |<
20^{\circ} $. This cut was suggested by \citet{PPMXLcorr}, who presented a
correction to the PPMXL proper motions. Although we do not make use of the
code provided to correct the proper motions, the Galactic latitude cut solves
the high discrepancy between the velocity fields of PPMXL and all other
catalogues. The applied cuts produce, respectively, samples of 173,162;
116,632 and 105,331 stars when UCAC4, PPMXL and Tycho\,-2 proper motions are
used. The adopted distance of each star is the inverse of the expectation of
the star's parallax, which \citet{binneydistances} recommended as the most
reliable distance estimate.

\subsection{The Tycho-Gaia astrometric solution catalogue} \label{sec:RAVE_TGAS}

An important aspect of RAVE DR5 is that with almost 256,000 spectra that overlap with a TGAS star,
this data release currently has a larger overlap with the Tycho-Gaia astrometric
solution (TGAS) catalogue than any other spectroscopic survey. TGAS is the
primary astrometric data set included in {\it Gaia} DR1, which includes
proper motions, sky positions and parallaxes of 2,057,050 stars from the
Hipparcos (\citealp{hipparcos}; \citealp{hipparcos2}) and Tycho\,-2
catalogues. In addition to the TGAS proper motions and positions, we use two
distance estimates for our RAVE-TGAS sample: the purely astrometric distances
of \citet{Astraatmadja} (ABJ), which employ an anisotropic prior based on a
three-dimensional density model of the Milky Way, and the improved RAVE
spectrophotometric distances from \cite{McMillan2017}, which take advantage
of both
TGAS parallaxes and effective temperatures from the Infrared Flux Method (IRFM;
\citealp{Blackwell1979}; \citealp{Casagrande2010}).

The RAVE-TGAS sample with McMillan distances involves a cut additional to
those listed in Section~\ref{sec:RAVE_data}: we removed stars with log\,$g<2$,
because distances to low-gravity stars are systematically overestimated (see \citealp{McMillan2017}). The data
selection results in a sample of 68,477 stars. 

For the RAVE-TGAS sample with ABJ distances we use the selection criteria
described in Section~\ref{sec:RAVE_data} less the cuts on heliocentric
velocity error, effective temperature, metallicity, surface gravity or the
convergence of the atmospheric parameters, because the ABJ distance estimates
do not rely on the accuracy of the atmospheric parameters. The RAVE-TGAS
sample with ABJ distances comprises 184,954 stars.

To further study the proper motion differences in the RAVE-TGAS samples, we
include the newly obtained UCAC5 \citep{UCAC5} proper motions, which were
derived by combining UCAC data with {\it Gaia} DR1.

\section{RAVE fifth data release}
\label{sec:DR5_wobbly}
\begin{table*}
	\begin{tabular}{c c c} 
		\\[6ex]
	\end{tabular}
	\caption{Largest gradient in $V_R$ obtained from the
		Williams/Kordopatis sample (W/K) and from the RAVE DR5 sample combined with
		three proper-motion catalogues. The values are computed using a least-squares
		fit (see Fig.\,\ref{fig:Gradients}). The uncertainties are taken from the
		covariance matrix.}
	\centering
	\begin{tabular}{c c c} 
		\\ 
		Catalogue& North $\partial V_R /\partial R$ [km\,s$^{-1}$ kpc$^{-1}$]& South $\partial V_R /\partial R$ [km\,s$^{-1}$ kpc$^{-1}$]\\ [1ex] 
		
		\hline\hline
		PPMXL& $-7.59\pm 1.15$&$-5.99\pm 0.81$\\ 
		Tycho\,-2& $-8.53\pm 1.82$&$-5.75\pm 0.47$ \\
		UCAC4& $-5.53\pm 1.24$&$-4.10\pm 0.89$ \\
		W/K& $-9.08\pm 1.48$&$-7.98\pm 0.87$ \\ [5ex]
		
	\end{tabular}
	\label{table:gradients}
\end{table*}

\citetalias{wobbly} studied the differences in the Galactocentric velocity
distributions for stars above and below the Galactic plane. For this, they
used a sample of 72,635 red-clump stars obtained from the RAVE internal third
data release \citep{DR3}. For comparison purposes, we use a similar data set
obtained from the internal fourth data release used in \citet{GK2013}. We apply the same cuts described in Section~\ref{sec:RAVE_data}. The data selection results in a sample of 127,722 stars (Williams/Kordopatis sample). The resulting velocity fields differ mainly from \citetalias{wobbly} in being
based on the full RAVE sample instead of only red-clump stars. This allows for a better
comparison to the velocity fields obtained with our RAVE samples.

In addition to the advantages of using RAVE DR5 data over RAVE DR4 (see Section~\ref{sec:RAVE_data}), a further improvement to previous works is the choice of proper motions used to compute the velocity fields. 
Indeed, \citetalias{wobbly} and \citet{GK2013} used an inhomogeneous compilation of proper motion values, a remnant of RAVE-DR3 approach; for a star present simultaneously in several proper motion catalogues available at the time, the proper motion with the smallest reported uncertainty was adopted. 
However, it is to be feared that combining proper motions measured with different techniques imprinted unknown systematics on the data set. Consequently, in this work, we do not mix the catalogues and use individual proper motions. The data presented here, with more homogeneous proper motions and better distance estimates, therefore yield velocities that are less systematically biased than the ones presented in \citetalias{wobbly} and \citet{GK2013}.

In Fig.\,\ref{fig:DR5_vphi} we present maps of the median azimuthal velocity
for each proper motion catalogue, as well as the one obtained by
Williams/Kordopatis. The velocity maps are displayed in (0.1\,kpc)$^2$ pixels
covering up to 3 kpc from the Sun. Each pixel is smoothed by computing the
median velocities over a box of (0.2\,kpc)$^2$ with a minimum of 50 stars.
The main characteristic observed in $V_{\phi}$ is the asymmetric drift.
Figure \ref{fig:DR5_vphi} displays a symmetrical distribution of velocity in
the northern and southern Galactic hemispheres, with velocities lagging that
of the LSR more at larger distance from the Galactic plane. In all
proper-motion samples, $V_{\phi}$ behaves similarly to the results obtained
by Williams/Kordopatis, which used a compilation of proper-motion catalogues:
within $0<\left | z \right |<0.5\,$kpc the median $V_{\phi}$ is in the order
of 220\,km\,s$^{-1}$ and decreases with distance from the plane to
190\,km\,s$^{-1}$ at $\left | z \right |\approx 1$ kpc and up to 130\,km\,s$^{-1}$ at
$\left | z \right |\approx 2$\,kpc.

The dependence of $V_R$ on proper motion is presented in
Fig.\,\ref{fig:DR5_vr}. Inwards of the solar radius, independent of the
proper motion catalogue used, stars both above and below the plane move away
from the Galactic centre with median velocities up to $V_R=30$\,km\,s$^{-1}$.
Below the plane in the range $-1<z<0$\,kpc, we estimate the largest radial
velocity gradient in the Williams/Kordopatis velocity field to be $\partial
V_R /\partial R = -7.98\pm 0.87$\,km\,s$^{-1}$\,kpc$^{-1}$, a value similar
to that found by \citet{siebertgradient}. Above the plane in the range
$0<z<1$\,kpc, $R
\lesssim 8$\,kpc, we estimate the largest value to be $\partial V_R
/\partial R = -9.08\pm 1.48$\,km\,s$^{-1}$\,kpc$^{-1}$. In Table
\ref{table:gradients} we report the largest $V_R$ gradients obtained from
each proper motion catalogue. The ($R,z$) planes used to obtain these
gradients are presented in Fig.\,\ref{fig:Gradients}.

The radial gradients in the northern and southern Galactic hemispheres are
consistent with one another within $1\sigma$. Thus, with these subsamples,
we find no North-South variation in the strength of the gradient within
8\,kpc. Consistent with the gradient reported by \citet{siebertgradient} and \citet{Anguiano2017}, the
gradients lie in the range $\sim 5-10$\,km\,s$^{-1}$\,kpc$^{-1}$.

Unlike $V_R$, $V_z$ depends largely on proper motions, which are vulnerable
to catalogue systematics and distance errors. Fig.\,\ref{fig:DR5_vz}
illustrates the strong dependence of the vertical velocity structure on the
proper-motion catalogue used. Inside the solar radius, all velocity fields
feature a positive stellar motion above the plane. The velocity fields
obtained with Tycho\,-2 and PPMXL show similar structure to that obtained by
Williams/Kordopatis: inside the solar radius we see upward stellar motion
above the plane and downward motion below it. By contrast, at $R<R_0$, the
UCAC4 velocity field shows upward motion both above and below the plane.
Hence inside the solar radius the PPMXL, Tycho\,-2 and Williams/Kordopatis
velocity fields have the signature of a breathing mode as discussed by
\citet{Widrow2012,Widrow2014}, while the UCAC4 shows that of a bending mode. 

Outside the solar radius, gradients in $V_z$ are weaker. PPMXL and Tycho\,2
show no clear signature, while UCAC4 and Williams/Kordopatis show weak
evidence for opposite patterns: in UCAC4 there appears to be upward motion
above the plane and downward motion below it, while the Williams/Kordopatis
field shows oppositely directed velocities. Either pattern would be
characteristic of a breathing mode.

Although the combination of breathing and bending modes outside and inside
$R_0$ shown by UCAC4 has not previously been reported in the literature, it
could arise from a combination of external and internal excitations. For
example, \citet{delaVega2015} used non-interacting test-particle integrations
to study the effects of a dwarf galaxy passage on the stellar epicyclic
motions and the resulting streaming caused by the subsequent phase-wrapping.
Since their simulations lacked of self-gravity, structures could not arise
from bending or breathing modes. Nonetheless, the velocity distribution
observed at an azimuthal angle $ \Theta=270^{\circ}$ in their Figure 8 is
consistent with the streaming motions caused by a combination of breathing
and bending mode.

However, the observed dependence of vertical velocity on proper motions
indicates the need for more accurate data. The ESA {\it Gaia} mission
is thus crucial in understanding the origins of the vertical streaming
motions observed in the Milky Way.

\section{Decomposition of the vertical velocity pattern}
\label{sec:Vz_pattern}

In this Section, before studying how the more precise {\it Gaia} data affects
the vertical velocity pattern as a whole, we study the composition of the
vertical velocity and the contribution of each component in the observed
structure. In \citet{1987} the Galactocentric vertical velocity $V_z$ is
obtained by computing the heliocentric velocity and adjusting for the solar
motion afterwards. However, by using the transverse velocity in Galactic
coordinates ($V_l$, $V_b$) and applying the solar motion correction before
computing $V_z$, we are able to separate the vertical velocity in two
components -- one involving the line-of-sight velocity corrected by the solar motion, $V_{\rm los}$, and the
other dependent on the distance, $d$, and the proper motion, $\mu$. The
Galactocentric vertical velocity is thus given by:
\begin{equation}
V_z=V_{\rm los}\,\sin b+V_{b}\,\cos b\\
\label{eq:vz}
\end{equation}
where
\begin{equation} V_b=4.74\,d\,\mu_b+(-U_{\odot}\,\cos l\,\sin b-V_{\odot}\,\sin l\,\sin b+W_{\odot}\,\cos b).
\label{eq:vb}
\end{equation}
Since the small uncertainty in $V_{\rm los}$ from RAVE does not affect the
$V_z$ pattern, we are in a position to study how distance and proper motion
affect the vertical velocity. First, we want to avoid the effects of disc
perturbations. For this we used Galaxia \citep{Galaxia} as a front end for
the (axisymmetric) Besan\c{c}on Galaxy Model \citep{Robin2003}.
Galaxia uses isochrones from the Padova database to compute photometric magnitudes for the model stars (\citealp{Bertelli1994}; \citealp{Marigo2008}) and it quickly generates from models mock observational catalogues that cover a specified area on the sky with any specified selection function. We use the Galaxia sample that was generated by \citet{Wojno2016} to obtain a mock stellar sample based on the RAVE selection function.

Figure \ref{fig:Galaxia} shows the median vertical velocity obtained from the
mock RAVE sample and its components. The top panels display $V_z$ considering
all components (left), only $V_{\rm los}\,\sin b$ (middle) and $V_{b}\,\cos
b$ (right, see Eq.\,\ref{eq:vz}). Here, we are able to identify, in dilute
form, the main patterns in $V_z$, in which at $R<8$ kpc $V_{\rm los}\,\sin b$ is
negative above the plane and positive below it, and at $R>8$\,kpc is
positive above the plane and negative below it. As expected, such a pattern
alone does not match any of the $V_z$ patterns found in
Fig.\,\ref{fig:DR5_vz}, indicating a stronger proper motion/distance
dependence. The top right panel of Fig.\,\ref{fig:Galaxia} does display key
aspects of the pattern visible in the W/K panel of Fig.\,\ref{fig:DR5_vz},
suggesting that in \citetalias{wobbly} errors may have caused the
proper-motion component of $V_z$ to dominate the more precise contribution
from $V_{\rm los}$. The $V_z$ subcomponents shown in the top middle and top
right panels have an inverse symmetry, which results in the structure-free
velocity field when summed up to estimate $V_z$ (top left panel).

In the bottom panels of Fig.\,\ref{fig:Galaxia}, we show how the distance,
proper motion and solar motion affect the transverse velocity component
$V_{b}\,\cos b$ (Eq.\,\ref{eq:vb}). In the bottom left panel we set $\mu_b=1$\,mas\,yr$^{-1}$
in the $V_{b}\,\cos b$ term and don't consider the solar motion. This displays the
effects of distance on $V_b$, where higher velocities are found at larger
distances. Similarly, in the bottom middle panel we set $d=1$\,kpc and exclude the
solar motion term, in order to see the effects of $\mu_b$ alone. Here, it is
shown that the proper motions contribute mainly with negative velocities
aside from the volume below the plane at $R>8$\,kpc. Finally, the right
bottom panel displays only the contribution to $V_b$ from the solar motion.
In all panels, the obtained structure is better understood by looking at Eq.\,\ref{eq:vz} and
Eq.\,\ref{eq:vb}. The term with $\cos b$ in the transverse velocity component
gives always a positive value, so the negative structure is given by $V_b$.
Here, the velocity pattern dependent on distance is also positive since
distances are inherently positive. In contrast, the proper motions and the
solar motion have negative values, with the solar motion remaining relatively
constant between samples. The different $V_b$ patterns are thus dependent
mainly on the proper motions, with the distance increasing its amplitude.
This can be seen in Fig.\,\ref{fig:GK_comp}, which displays the components of
$V_z$ in the Williams/Kordopatis sample. Here, although the pattern obtained
from the distance has lower velocities than in Galaxia, the pattern of proper
motions is strong enough to dominate the contribution from $V_{\rm los}$,
breaking the balance previously shown in Galaxia and generating the pattern
of a breathing mode.

In Fig.\,\ref{fig:Galaxia_errors} we study the effects of introducing errors
to the mock data from the axisymmetric and relaxed Galaxia model. The first panel
shows the Galaxia sample with no errors. In the second panel we introduce
artificial errors and draw new velocity values, from a normal distribution
centred on the actual velocity and $\sigma_{\mu}=5$\,mas\,yr$^{-1}$. Although
these errors are larger than the typical error in our DR5 samples (up to
$\left \langle \sigma_{\mu}\right \rangle\sim4$\,mas\,yr$^{-1}$), the
assigned values produce some clumps but do not affect the general structure
of the sample.

The third panel shows the effect of adding a random distance error with a
standard deviation $\sigma_d/d=0.8$ (we chose such large errors to show
a clear pattern, although this is already seen with $\sigma_d/d=0.5$). This
random error increases the volume of our sample and, since all distance values are
positive, increases the value of the $V_b$ component (Eq.\,\ref{eq:vb}). This
results in a pattern similar to the one obtained by Williams/Kordopatis. The
right panel displays a stronger more noisy pattern as a combination of both
random errors. The obtained patterns in this Section allow us to understand
better how each component affects $V_z$. In Section~\ref{sec:RAVE-TGAS} we
will analyse how this compares to our RAVE-TGAS sample.

\section{RAVE-TGAS sample} \label{sec:RAVE-TGAS} 

The superior parallaxes and proper motions obtained by the ESA {\it Gaia}
mission should yield marked improvements on the 
velocity fields presented in Section~\ref{sec:DR5_wobbly}.
In Section~\ref{sec:TGAS_Distance} we first compare the distance estimates
obtained purely from {\it Gaia} DR1 with the improved RAVE DR5
spectrophotometric distances. Then in Section~\ref{sec:TGAS_PMs} we
compare the impact of using proper motions from RAVE DR5, TGAS and UCAC5. In
Section~\ref{sec:TGAS_wobbly} we introduce the velocity fields obtained using
the best estimates in distance and proper motion available for our RAVE-TGAS
sample.

\subsection{Distance estimate} \label{sec:TGAS_Distance} 

As discussed in Section~\ref{sec:RAVE_TGAS}, our RAVE-TGAS sample makes use
of two distance estimates: those of \citet{Astraatmadja} and
\cite{McMillan2017}. We do this on account of the problems discussed by
\citet{CBJ1} associated with use of inverse parallax as a distance estimator.

\begin{figure}
\includegraphics[width=\columnwidth]{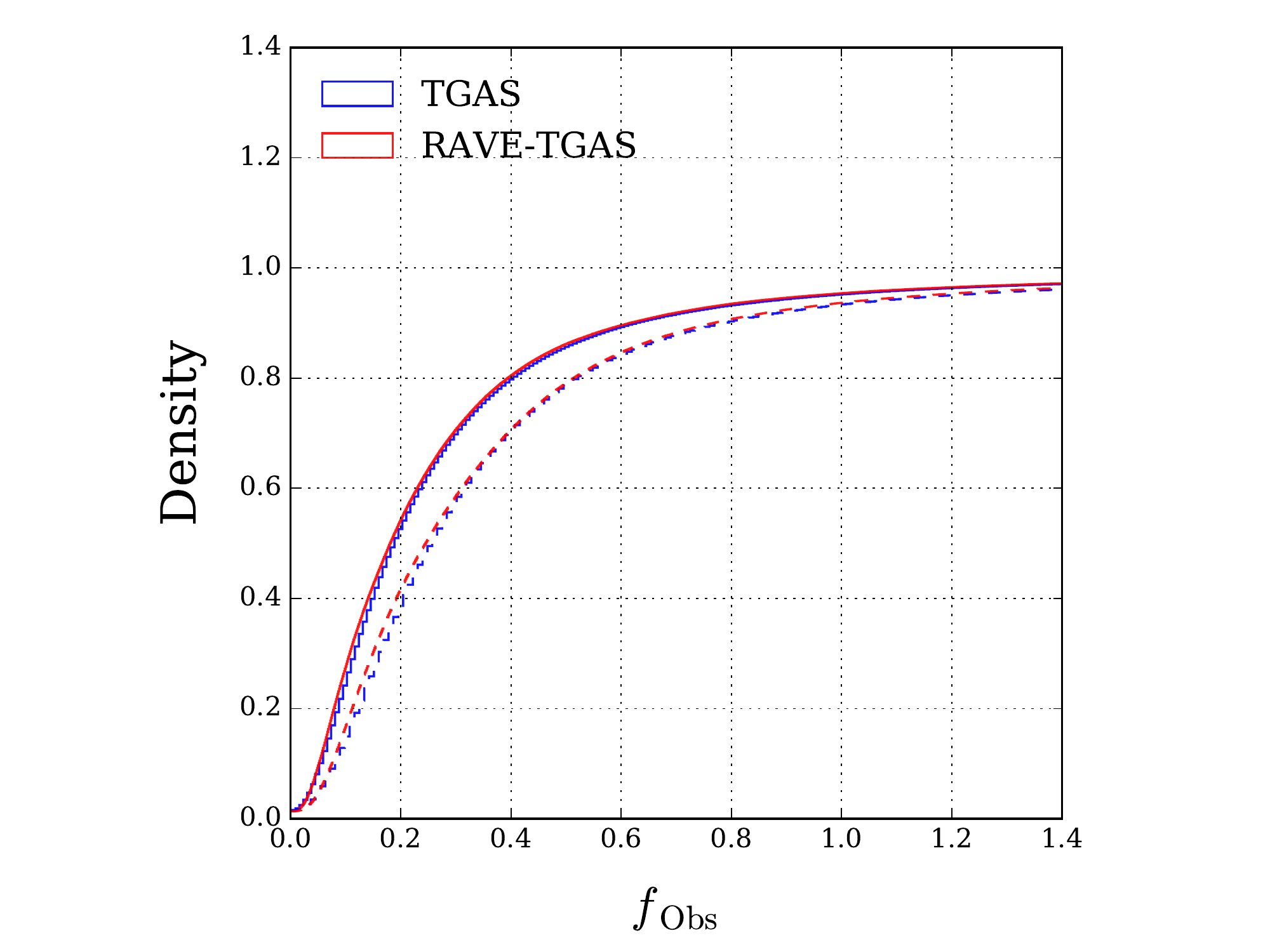}
\caption{Cumulative distribution of the fractional parallax errors $f_{\rm Obs}$
of stars in the full TGAS (blue line) and RAVE-TGAS (red line) samples. The
dotted curves represent the same samples taking into account systematic errors in the parallax uncertainties. Although there is a
significant number of stars with $f_{\rm Obs}<0.2$, using the inverse parallax as
a distance estimate produces biased distances for at least $\sim50\%$ of the
stars in both samples.} \label{fig:fobs}
\end{figure}

If we assume measured parallaxes $\varpi$ to be normally distributed with a standard
deviation $\sigma_{\varpi}$, and adopt the inverse parallax as the distance
estimator, we encounter two important issues: (i) the estimator fails for
negative $\varpi$, even though these are valid measurements (see
\citealp{CBJ1} for further details). (ii) For fractional parallax errors
$f_{\rm Obs}=\sigma_{\varpi}/\varpi>0.2$, using the inverse parallax creates a
skewed distribution which gives a biased distance estimator. Figure
\ref{fig:fobs} shows a cumulative histogram of $f_{\rm Obs}$ for our RAVE-TGAS
sample as well as the full TGAS sample. The parallaxes of {\it Gaia} DR1
suffer from a possible offset by $\pm0.1$ mas of the parallax zero point and
$\pm0.2$ mas due to position and colour-dependent systematics. A detailed
description of these systematics is given in \citet{GaiaDR1}, appendix E. It
is therefore recommended to add a systematic error of $\pm0.3$\,mas to
the parallax uncertainties \citep{GaiaDR1first}. The dotted curves show the
same TGAS data but consider the systematic errors.
As can be seen, the full TGAS and our RAVE-TGAS sample have a relatively
large number of stars ($\sim50\%$) within $f_{\rm Obs}<0.2$. Nevertheless, using
the inverse parallax in the RAVE-TGAS sample would lead to potentially biased
distance estimates for a minimum of $\sim50\%$ of the stars. If we consider
the systematic errors, we see that the distance estimate would be
biased for at least $\sim60\%$ of the stars.

The inverse parallax is therefore a poor distance estimate for our RAVE-TGAS
sample. If we select only the stars with $f_{\rm Obs}<0.2$ including systematic
errors, we would limit the observed volume of stars within just 0.5\,kpc
around the Sun. The volume obtained is then not big enough to recognize
a distinctive mode in the Galaxy. Hence, another approach to estimate
distances is required. Recently \citet{Astraatmadja} used Bayesian inference
to estimate distances for the full TGAS sample. For this, they used two
different priors: a exponentially decreasing space density prior and an
anisotropic prior based on a three-dimensional density model of the Milky
Way. The consistency of both priors was tested by comparing their distance
estimates with the determined distances of 105 Cepheid variable stars from
\citet{Cepheids} cross-matched with {\it Gaia} DR1 (see Figure 4 of
\citet{Astraatmadja}). It was shown that for distances $d<2$\,kpc, the Milky
Way prior comes closer to the more precise Cepheid distance estimate than the
exponentially decreasing space density prior. Beyond this limit, the latter
performs better. Due to the volume of our samples ($d\lesssim 2$\,kpc), we choose
the distance estimate obtained from the Milky Way prior for further analysis
in this work.

Recently, \cite{McMillan2017} determined new distances to RAVE stars also
using Bayesian inference, but considering both TGAS parallaxes and stellar
parameters derived from the RAVE spectra, with $T_{\mathrm{eff}}$ values
derived using the Infrared Flux Method (IRFM). Therefore, it is recommended
to use this new estimate when combining RAVE and TGAS data.

\begin{figure}
\includegraphics[scale=.57]{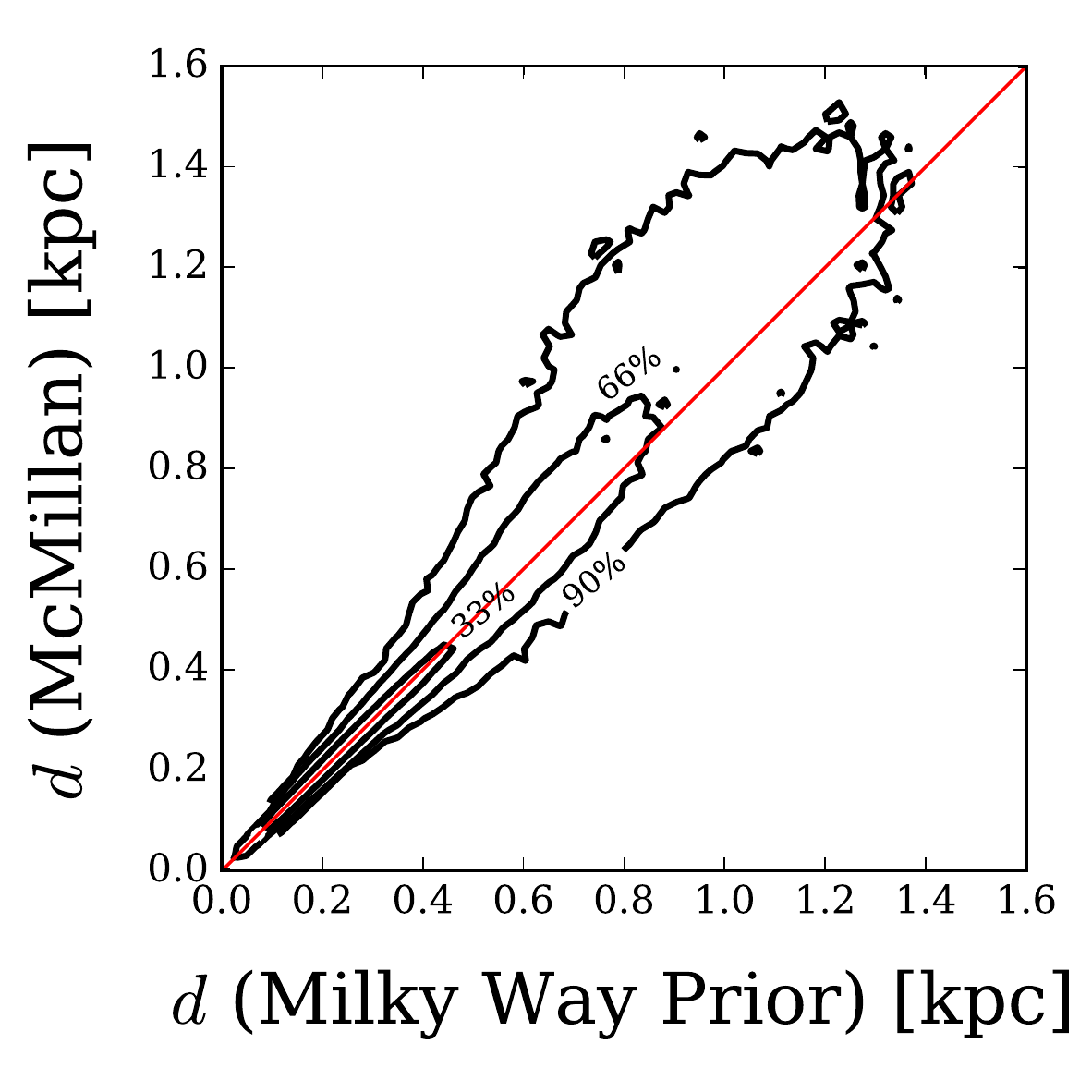} \caption{Comparison of the distance
estimates of McMillan, which combine TGAS parallaxes with spectrophotometric
data, and those of ABJ using the Milky Way prior. The red line indicates
a perfect match between distances. Under 1\,kpc about $\sim66\%$ of the
distance estimates are in good agreement.}
\label{fig:ABJdist_RAVE}
\end{figure}

\begin{figure}
	\includegraphics[scale=.57]{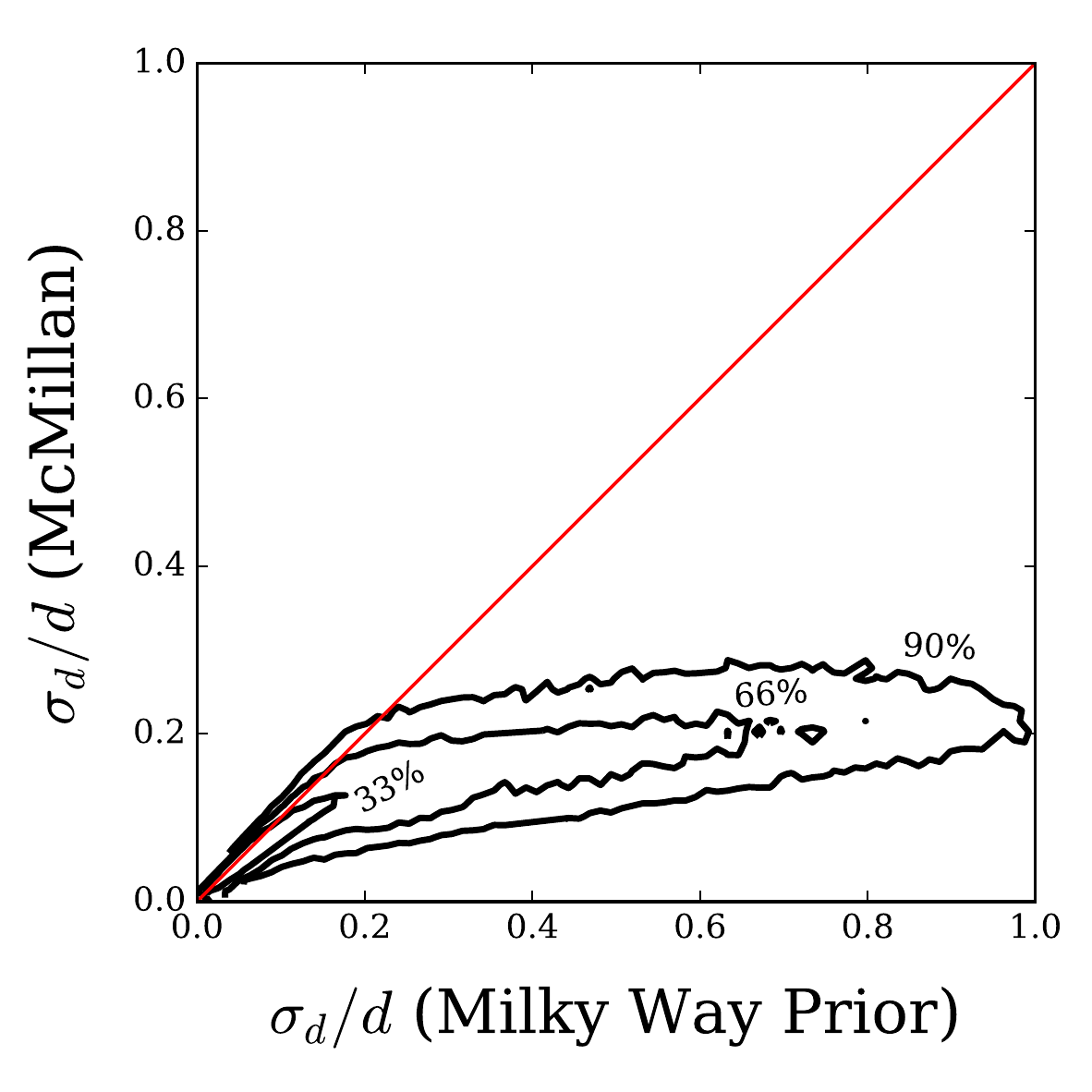} \caption{Comparison
		between relative uncertainties $\sigma_d/d$ in the distance estimates of
		McMillan and ABJ. The red line denotes equal $\sigma_d/d$ values. For most
		stars the McMillan distances have the smaller relative uncertainties.}
	\label{fig:ABJ_RAVE_relerr}
\end{figure}

Figure \ref{fig:ABJdist_RAVE} presents a comparison between the McMillan distances and the ABJ distance estimates with the Milky Way prior.
The residuals between the two distance estimates have a mean value of
$\mu=0.10$\,kpc and a dispersion of $\sigma=0.36$\,kpc. Within $\sim1$\,kpc
the agreement is very good. A comparison of the relative uncertainties $\sigma_d/d$ between both estimates is presented in Fig.\,\ref{fig:ABJ_RAVE_relerr}.
\begin{figure*}
	\includegraphics[width=\textwidth]{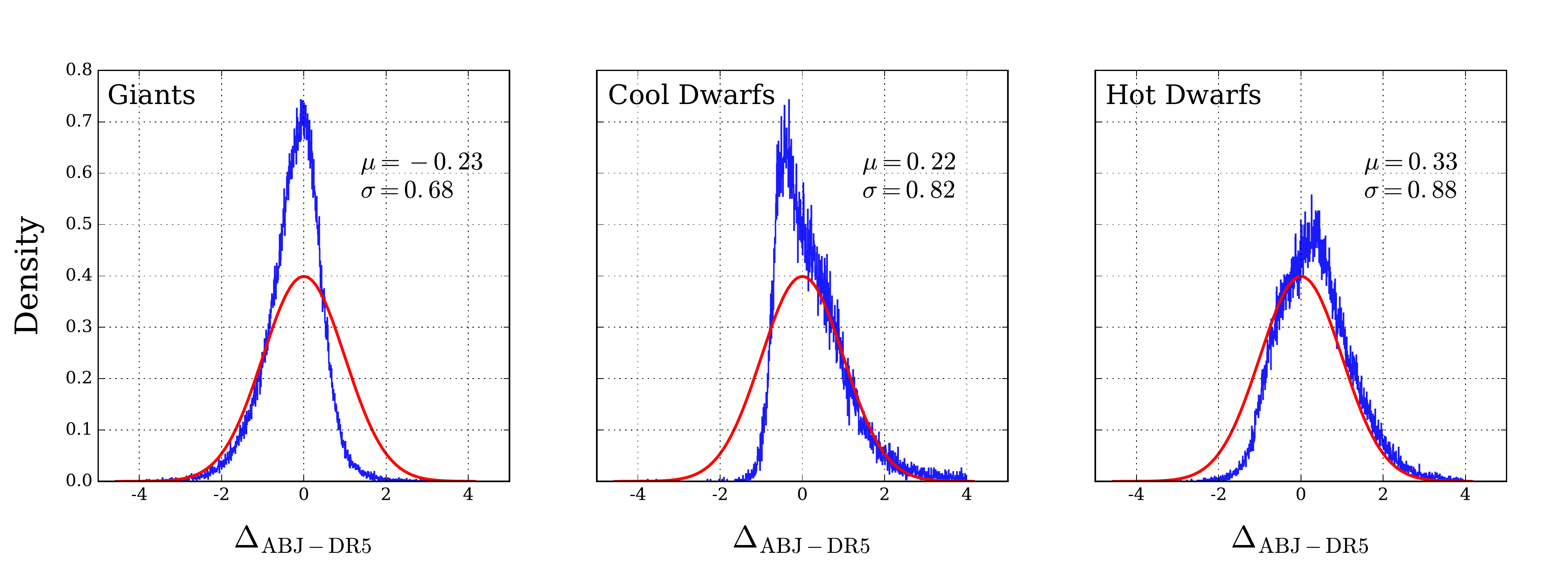}
	\caption{Histograms of the
		difference between the distance of ABJ and RAVE DR5. Stars
		are divided into giants (log\,$g<3.5$), hot dwarfs (log\,$g>3.5$ and
		$T_{\mathrm{eff}}>5500$\,K) and cool dwarfs (log\,$g>3.5$ and
		$T_{\mathrm{eff}}<5500$\,K). The differences are computed using
		Eq.\,\ref{eq:Delta}, the solid red curves are Gaussians of zero mean and unit
		dispersion, representing perfect results. To remove outliers, we use only values with $\left |\Delta_{\rm ABJ-DR5}  \right |< 4$. The most characteristic feature in all three panels is the non-zero mean, indicating a clear difference between both distance estimates and a dispersion less than unity, corresponding to overestimated uncertainties of one or both measurements.} \label{fig:ABJdist_RAVE_hist}
\end{figure*}
As can be seen, most stars have smaller relative uncertainties when using the
McMillan estimate.  Due to the consistency between the distance estimates we want to further study the ABJ distance uncertainties. Since the McMillan and the ABJ distance estimates are not totally independent -- both estimates use TGAS parallaxes to infer distances, we are not able to compare their differences and learn about biases in one measurement versus another or whether the error estimates are accurate. Therefore, we compare ABJ to RAVE DR5 distances to further study the ABJ distance uncertainties.

Here, as in \citet{binneydistances} and \citet{RAVEDR5} we divide the sample in three groups of stars: giants (log\,$g<3.5$), hot dwarfs (log\,$g>3.5$ and $T_{\mathrm{eff}}>5500$\,K) and  cool dwarfs (log\,$g>3.5$ and $T_{\mathrm{eff}}<5500$\,K). The histograms of Fig.\,\ref{fig:ABJdist_RAVE_hist} show the differences between the ABJ distance and the RAVE DR5 distances as follows:
\begin{equation}
\Delta_{\rm ABJ-DR5} =\frac{d_{\rm ABJ}-d_{\rm
DR5}}{\sqrt{\sigma_{d,\rm ABJ}^{2}+\sigma_{d,\rm DR5}^{2}}}
\label{eq:Delta}
\end{equation}
Ideally, $\Delta_{\rm ABJ-DR5}$ would have zero mean (no biases
in one measurement versus another) and unit dispersion (consistent with
the uncertainties being correctly estimated). The solid red curves are
Gaussians with the desired mean and dispersion. For hot and cool dwarfs the mean value of $\Delta_{\rm ABJ-DR5}$ is positive and deviates from zero, which shows that the distance bias is a significant fraction of the uncertainty, with the ABJ distances being larger than the RAVE DR5 distances. The negative $\Delta_{\rm ABJ-DR5}=-0.23$ for Giants indicate larger distances in DR5. The characteristic feature in all three panels is the dispersion $\sigma<1$, this implies that the uncertainties of one or both measurements are overestimated. However, the agreement of distance estimates in Fig.\,\ref{fig:ABJdist_RAVE} combined with the differences between relative uncertainties in Fig.\,\ref{fig:ABJ_RAVE_relerr} suggest that the ABJ distance uncertainties are overestimated. This could be due to the possible overestimation of TGAS uncertainties as discussed in \cite{McMillan2017}.
\subsection{TGAS proper motions}
\label{sec:TGAS_PMs} 

\begin{figure}
\includegraphics[scale=.48]{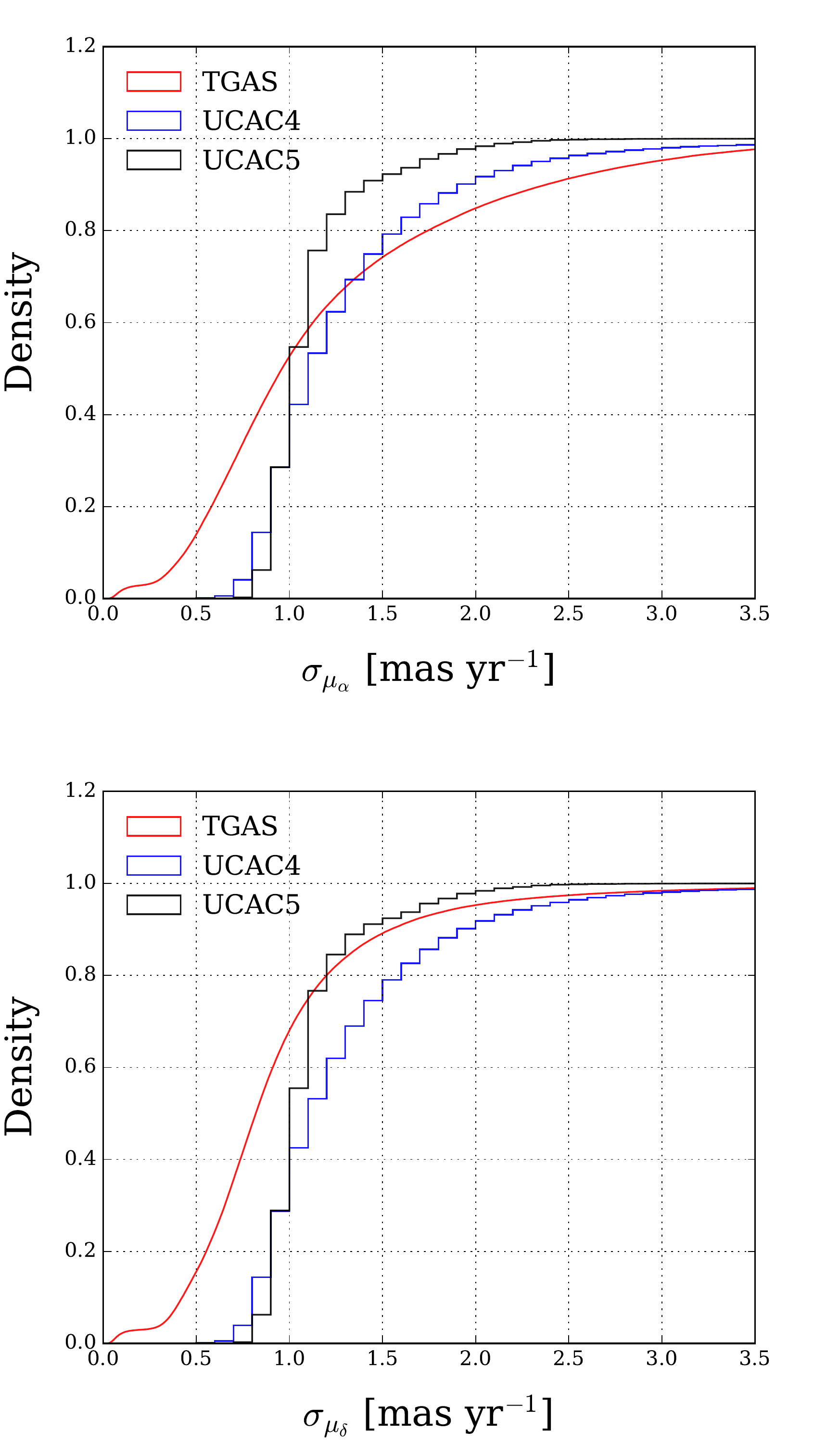}
 \caption{Cumulative distribution of uncertainties in proper-motion in right
ascension $\sigma_{\mu_{\alpha}}$ (top) and in declination $
\sigma_{\mu_{\delta }}$ (bottom). The blue line is for UCAC4, the black line
is for UCAC5, and the red line is for TGAS. Up to $\sim70\%$ of TGAS proper
motions have uncertainties smaller than 1 mas\, yr$^{-1}$. In UCAC5 and UCAC4 the same
is true only for very few stars.}
\label{fig:TGAS_Pms_comparesigma}
\end{figure}

\begin{figure}
\includegraphics[width=\columnwidth]{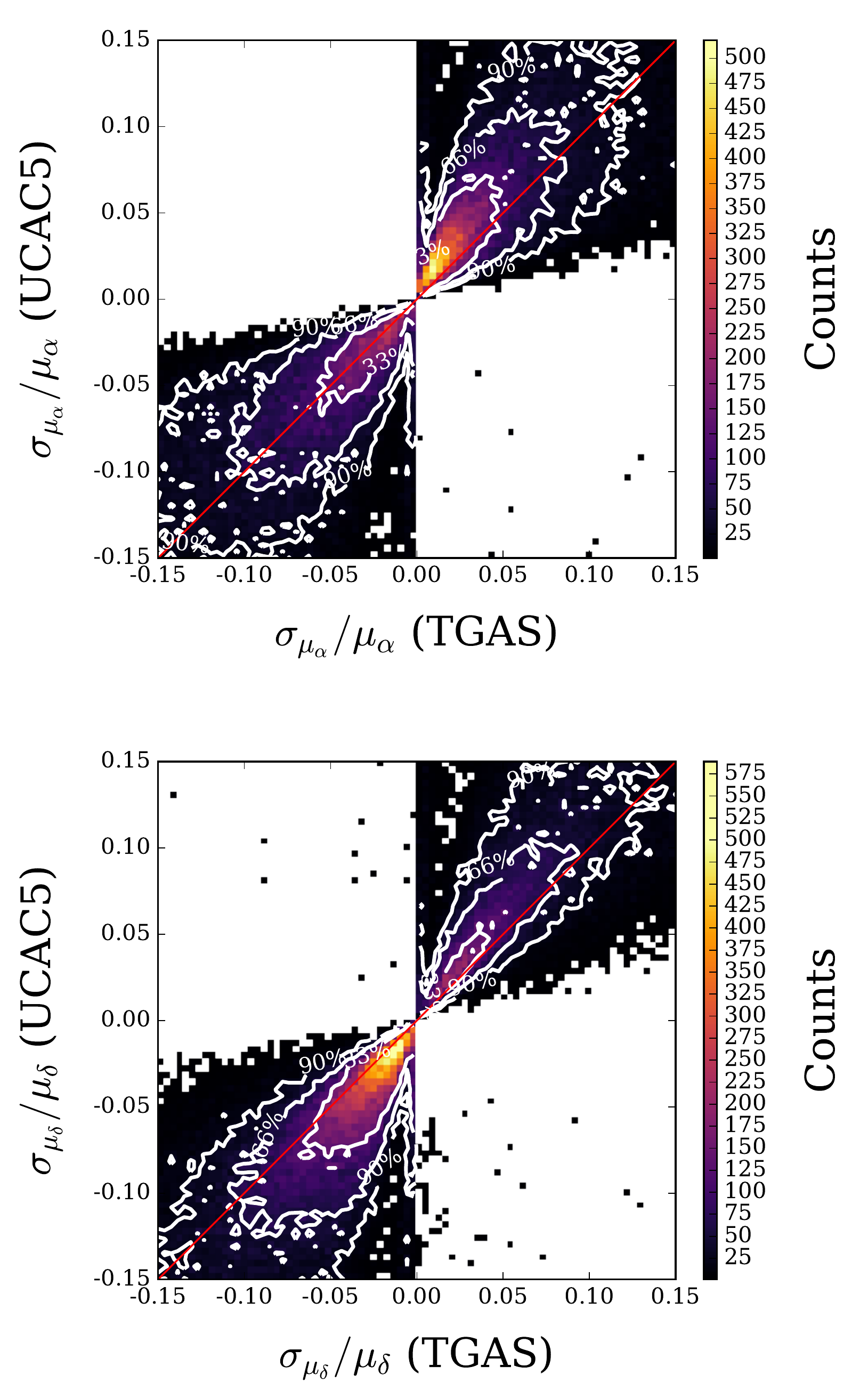} 
 \caption{Comparison between relative uncertainties in proper motion in right
ascension $\sigma_{\mu_{\alpha}}/\mu_{\alpha}$ (top) and declination $
\sigma_{\mu_{\delta }}/\mu_{\delta }$ (bottom) in the UCAC5 and TGAS
catalogues. In both panels, independent of direction, most of the stars have
larger relative uncertainties in UCAC5 than in TGAS.}
\label{fig:TGAS_Pms_compare}
\end{figure}

In RAVE the UCAC4 catalogue is the last proper-motion catalogue that is independent
of {\it Gaia}, and it contains the largest number of stars in RAVE DR5. So we use
the RAVE-TGAS sample to compare UCAC4 proper motions with TGAS and UCAC5
proper motions. Figure \ref{fig:TGAS_Pms_comparesigma} displays the
cumulative distribution of the uncertainties in proper motion. The upper
panel is for the right-ascension component, $\sigma_{\mu_{\alpha}}$, while
the lower panel is for the declination component, $ \sigma_{\mu_{\delta }}$.
In UCAC4 and UCAC5 very few stars have uncertainties smaller than
1\,mas\,yr$^{-1}$, whereas in TGAS more than $60\%$ of stars have smaller
uncertainties in declination and only slightly fewer have smaller
uncertainties in right ascension. On the other hand, less than 30\% of stars
have UCAC5 uncertainties larger than 1\,mas\,yr$^{-1}$ in either component
while both TGAS and UCAC4 have longer tails of stars with larger
uncertainties in right ascension. Since our sample has a mean/median
$\sigma_{\mu}\sim 1 $\,mas\,yr$^{-1}$, the TGAS proper motions are more
reliable in the selected volume.
The improvement in precision provided by TGAS proper motions rather than
UCAC5 is shown in Fig.\,\ref{fig:TGAS_Pms_compare}.
Both panels indicate smaller relative proper-motion uncertainties in TGAS
than in UCAC5 in that the ridge-lines of the distributions are steeper than
the red lines for a perfect match between catalogue values. The mean
relative uncertainties in right ascension for the most precisely observed
90$\%$ of the stars are of order 6.0$\%$ in UCAC5 and 4.3$\%$ in TGAS for
positive proper motions and 6.6$\%$ in UCAC5 and 6.3$\%$ in TGAS for negative
proper motions.

Since even small proper motions differences can significantly affect the
space velocities obtained, the use of TGAS/UCAC5 data over UCAC4 takes us one step
closer to solving the Milky Way mode discrepancy.

\subsection{Gaia's wobbly Galaxy} \label{sec:TGAS_wobbly} 

\begin{figure*}
\includegraphics[width=\textwidth]{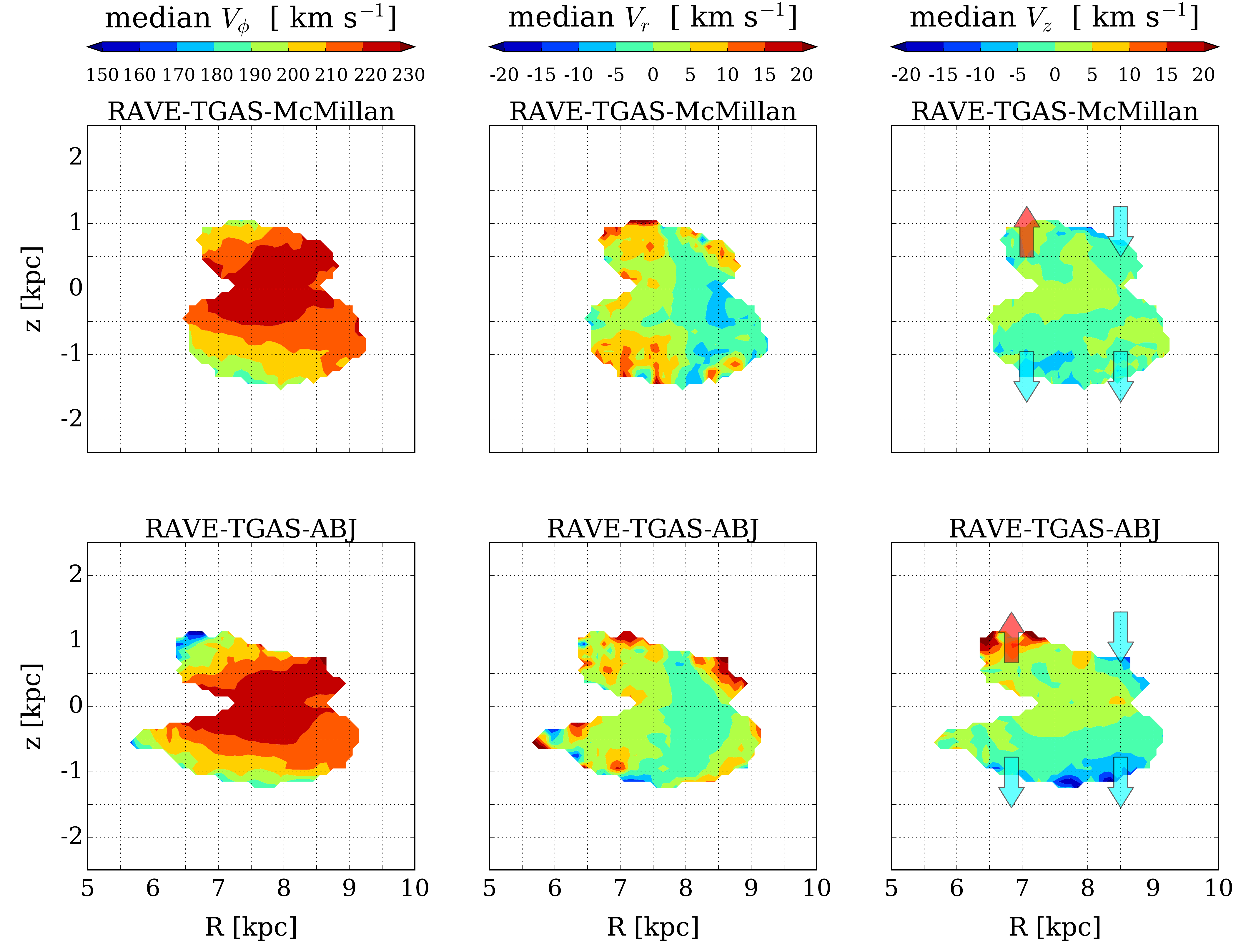}
 \caption{Maps of median values of each component of Galactocentric velocity.
 All panels employ RAVE line-of-sight velocities and TGAS proper motions.
The top panels employ \citet{McMillan2017} distances with $\sigma_d/d<25\%$,
while the bottom panels employ the distances derived by \citet{Astraatmadja}
using the Milky Way prior. In contrast to Fig.\,\ref{fig:DR5_vr}, the middle panels for $V_R$ display a velocity gradient above the plane. The right panels for $V_z$ show the
signatures of a breathing mode perturbation inside and
a bending mode outside $R_0$.} \label{fig:RAVE_TGAS_ABJ}
\end{figure*}

We now present the velocity fields constructed from the most accurate
estimates for our sample.
When using the McMillan distances, we exclude stars with an
error of $\sigma_d/d>25\%$ to exclude uncertain velocities from our velocity
fields. We join these distances to line-of-sight velocities from RAVE and
TGAS proper motions to form the RAVE-TGAS-McMillan data set with 58,972 stars.
The top panels of Fig.\,\ref{fig:RAVE_TGAS_ABJ} show the median values of
$V_{\phi}$, $V_R$ and $V_z$ as a function of galactic position for
the RAVE-TGAS-McMillan set. The bottom panels show the corresponding velocity
fields for the RAVE-TGAS-ABJ set. Since the ABJ distance uncertainties appear
to be overestimated (Section~\ref{sec:TGAS_Distance}), no distance cut has
been applied.

Both the $V_{\phi}$ velocity fields shown in the left panels of
Fig.\,\ref{fig:RAVE_TGAS_ABJ} agree with those computed in
Fig.\,\ref{fig:DR5_vphi} using other data sets and proper motions. Within
500\,pc, stars both above and below the disc midplane exhibit velocities $V_{\phi}\sim 220$\,km\,s$^{-1}$ which decrease with $z$. At $z\sim1$\,kpc
and despite the smaller observed volume, the azimuthal velocity in the bottom
left panel reaches $V_{\phi} \sim 170$ km\,s$^{-1}$, similar to the values
observed with Tycho\,-2 in Fig.\,\ref{fig:DR5_vphi}.

In the middle panels of Fig.\,\ref{fig:RAVE_TGAS_ABJ}, both maps of median
$V_R$ differ from the maps obtained with the other proper
motion catalogues (see Fig.\,\ref{fig:DR5_vr}). At $0.5<z<1$\,kpc a negative
radial velocity gradient is evident above as well as below the midplane.
When the McMillan distance is used, the peak gradient below the plane is
$\partial V_R /\partial R= -7.01 \pm 0.61$\,km\,s$^{-1}$\,kpc$^{-1}$,
consistent with previous values, while above the plane the peak gradient is
$\partial V_R /\partial R= -9.42 \pm 1.77$\,km\,s$^{-1}$\,kpc$^{-1}$. The
ABJ distances yield similar values: $\partial V_R /\partial R= -6.01 \pm
0.62$\,km\,s$^{-1}$\,kpc$^{-1}$ below the plane and $\partial V_R /\partial
R= -9.24 \pm 2.28$\,km\,s$^{-1}$\,kpc$^{-1}$ above it. Both maps show similar
radial-velocity structure, with a negative gradient in both hemispheres
inside $R_0$ and a positive gradient at $R>R_0$. This radial structure could
be related to the simulations of \citet{2014Faure} and the analytical results
of \citet{2016Monari}, who studied the response of stars to a stable spiral
perturbation. They found that the mean value of $V_R$ is negative within the arms and
positive in the interarm region.

\begin{figure}
	\includegraphics[width=\columnwidth]{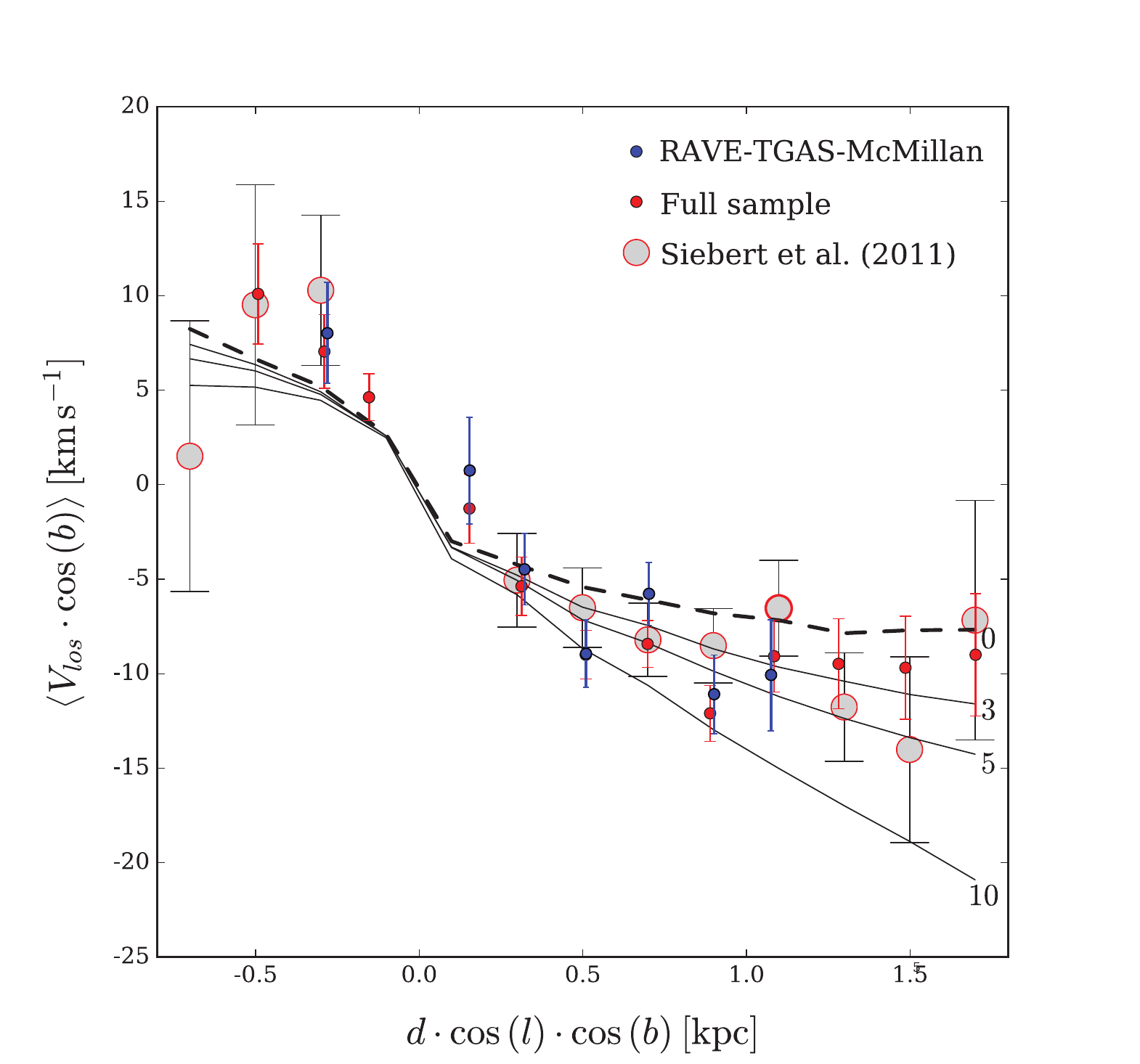}
	\caption{Mean of the component of line-of-sight velocity within the
	plane for stars that lie close to the direction to either the
	Galactic centre ($|l|<5^\circ$) or anticentre
	($175^{\circ}<l<185^{\circ}$) versus projected position along the
	centre--anticentre line. The mean is computed in bins of 0.2\,kpc
with a minimum of 50 stars per bin from the RAVE-TGAS-McMillan sample and the
RAVE DR5 full sample compared to the results obtained by
\citet{siebertgradient}. The full curves represent a thin disc with the
indicated radial-velocity gradient, while a thick disc with zero motion is
represented by the dashed curve. All three samples agree within their
uncertainties.}
	\label{fig:TGAS_Siebert}
\end{figure}

As a comparison to the radial gradient reported by \cite{siebertgradient},
Fig.\,\ref{fig:TGAS_Siebert} shows the projection of line-of-sight velocity
in the direction of the Galactic centre ($\left | l \right |<5^{\circ}$) and
anticentre ($175^{\circ}<l<185^{\circ}$) for the RAVE-TGAS-McMillan sample,
the RAVE DR5 full sample and the results obtained in their Figure 3. We
compute the mean line-of-sight velocity in bins of 0.2\,kpc with a minimum of
50 stars per bin. Each full curve represents the expected velocities
of a thin-disc model with the radial velocity gradients as indicated. The
dashed curve represents a thick-disc model with zero gradient (see
\citealp{siebertgradient} for further details). As can be seen both our
samples are consistent within $1\sigma$ with their results.

The right panels of Fig.\,\ref{fig:RAVE_TGAS_ABJ} display maps of median
vertical velocities. Inside $R_0$ we see a breathing mode with positive
velocities above the plane and negative below; however outside $R_0$, we see
negative velocities above and below the plane, corresponding to a bending
mode. This combination of breathing and bending modes is inverted relative to
the combination discussed in Section~\ref{sec:DR5_wobbly} on the basis of the
UCAC4 proper motions, and it also differs from the pattern found by
\citetalias{wobbly}. The mode change we observe at $R \approx R_0$ could be attributed to the Galactic warp as mentioned before, but a deeper analysis is required to understand this effect. By comparing both right panels of Fig.\,\ref{fig:RAVE_TGAS_ABJ} with the pattern obtained by Williams/Kordopatis,  it is clear that in both our samples $V_{\rm los}$ and $V_b$ contribute equally, whereas the $V_z$ pattern obtained by Williams/Kordopatis is consistent with a dominant $V_b$ contribution (see Fig.\,\ref{fig:GK_comp}).

\begin{figure*}
	\includegraphics[scale=0.53]{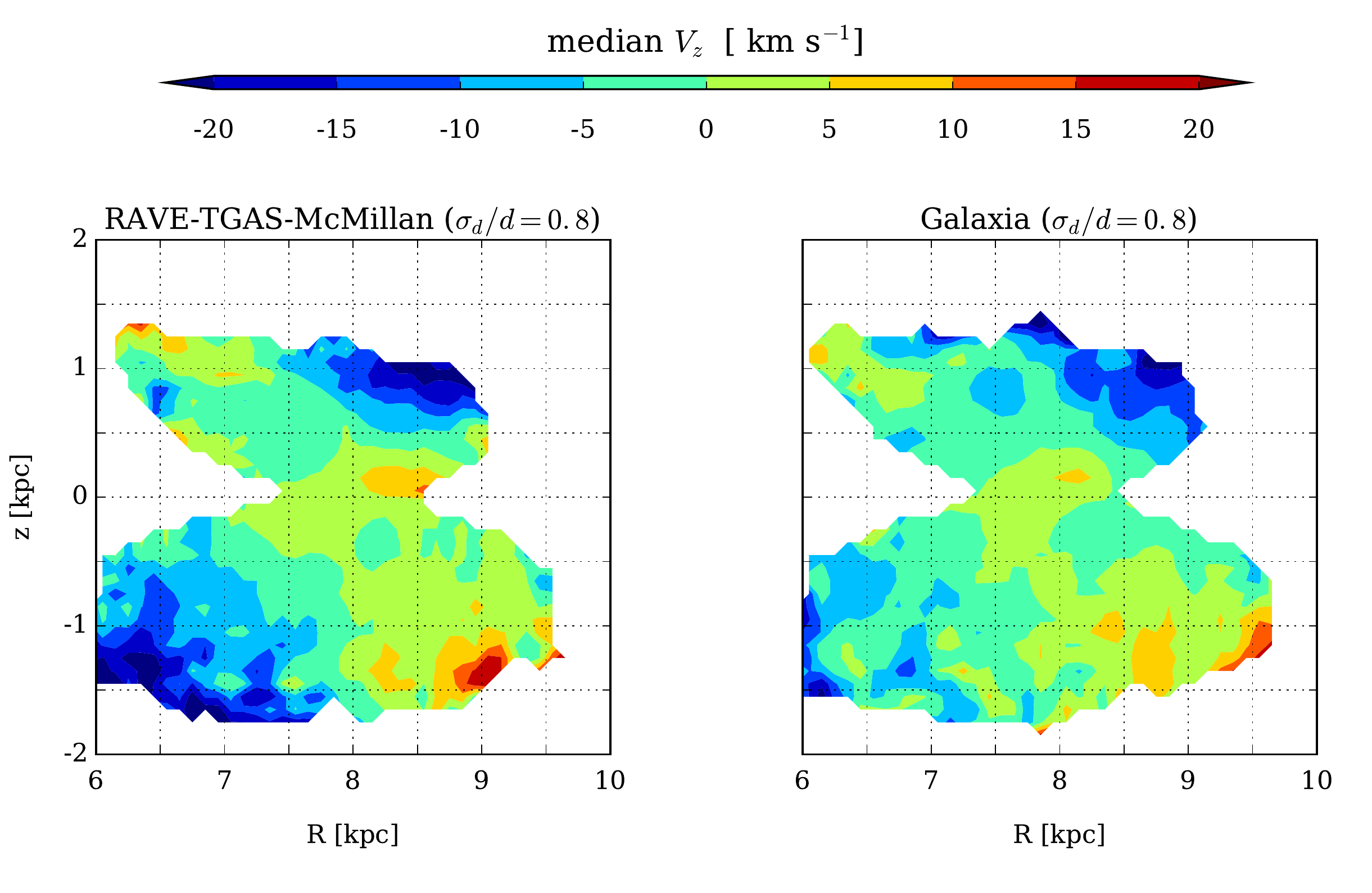}
	\caption{Comparison of $V_z$ from the RAVE-TGAS-McMillan
		sample (left) and the RAVE-like sample from Galaxia (right). Similar to Fig.\,\ref{fig:Galaxia_errors}, the  panels display the $V_z$ pattern obtained by adding a random error in distance with dispersion $\sigma_d/d=0.8$. The result is a
		pattern similar to the one obtained with the Williams/Kordopatis sample.}
	\label{fig:TGAS_Galaxia}
\end{figure*}

As in Section~\ref{sec:Vz_pattern}, we now study the effects of introducing errors to our sample and compare these to the results obtained with Galaxia using the same data selection. Figure \ref{fig:TGAS_Galaxia} shows this comparison. Although the RAVE-TGAS-Mcmillan sample exhibits vertical structure and Galaxia is axisymmetric, by adding a random error in distance of $\sigma_d/d=0.8$, we increase the dependence of $V_z$ on $V_b$ and thus reproduce a velocity pattern similar to that found by Williams/Kordopatis. This experiment suggests
that the different relative importance between the $V_{\rm los}$ and $V_b$ patterns
in the RAVE-TGAS-McMillan sample and the ones obtained by Williams/Kordopatis
could arise from better distance estimates for our sample.

To study further the combination of breathing and bending modes, we now divide
our subsamples in different radial bins from $R=6$\,kpc to $R=9$\,kpc in
0.5\,kpc intervals and compute the median $V_z$ as a function of $z$ in bins
of 0.3\,kpc. Each bin contains a minimum of 50 stars.

\begin{figure*}
	\includegraphics[scale=.40]{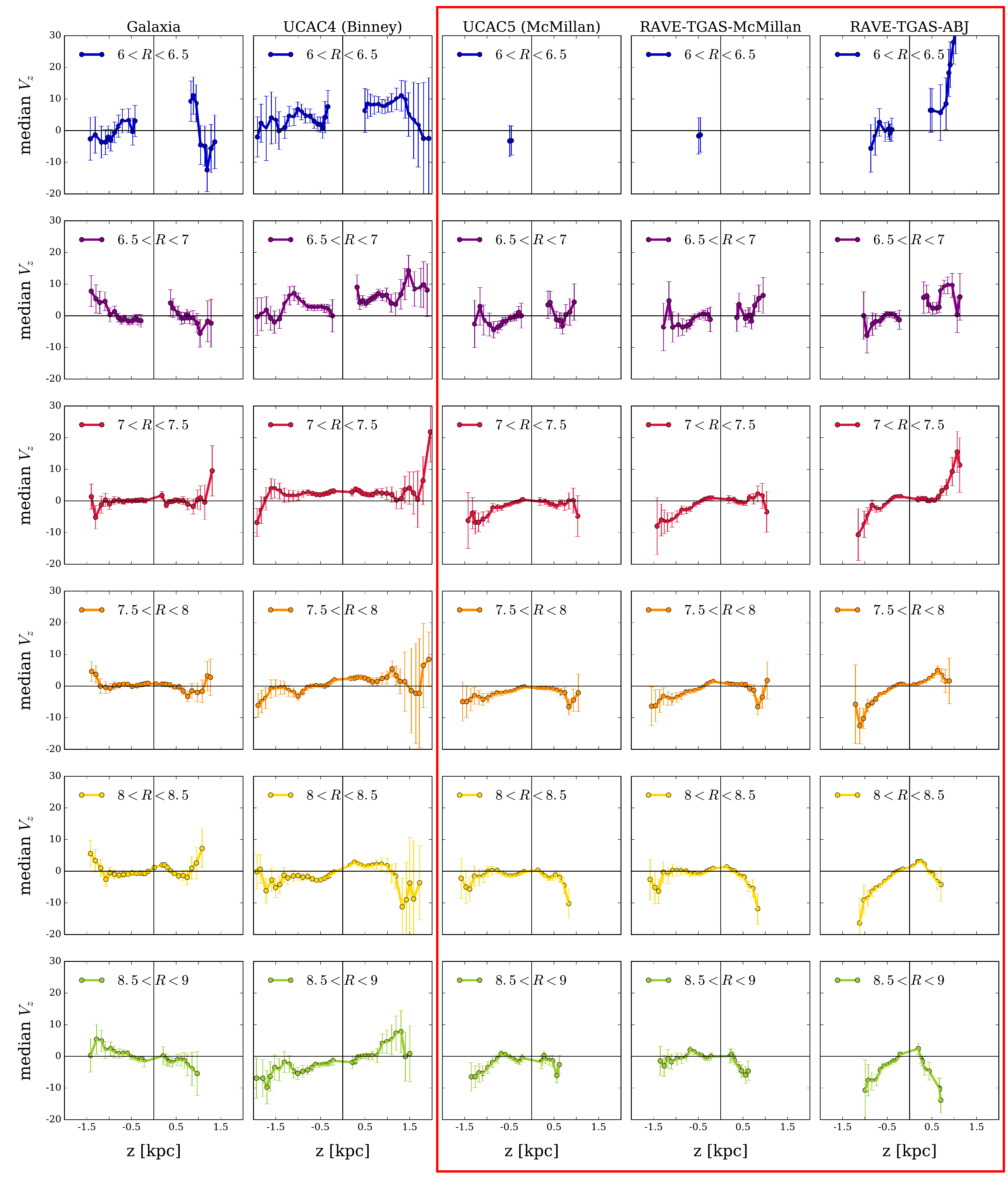}
	\caption{Median values of $V_z$ as a function of $z$ for different $R$
		ranges for the axisymmetric Galaxia model with random errors of $\sigma_d/d=0.8$ and $\sigma_{\mu_b}=1$\,mas\,yr$^{-1}$ (1\textsuperscript{st} column) and using proper motions from:
		UCAC4 with RAVE DR5 distances (2\textsuperscript{nd} column); UCAC5 with McMillan distances cut on
		$\sigma_d/d<25\%$ (3\textsuperscript{rd} column), TGAS with McMillan
		distances cut on $\sigma_d/d<25\%$ (4\textsuperscript{th} column); and TGAS
		with ABJ distances (5\textsuperscript{th} column). Overlapped bins are
		computed every 0.3\,kpc with a minimum of 50 stars per bin. The error bars
		correspond to the standard error of the median. The upper panels in the three
		right columns (which all benefit from TGAS proper motions) exhibit far from the plane
		positive velocities in the northern hemisphere and negative in the southern
		hemisphere, the signature of a breathing mode. The 
		lower panels in these columns show the signature of a bending mode, even parity in
		the vertical velocity distribution. By contrast, the lower panels for UCAC4
		show the signature of a breathing
		mode.} \label{fig:Widrowcompare}
\end{figure*}

From left to right Fig.\,\ref{fig:Widrowcompare} shows the results obtained
with (i) axisymmetric Galaxia model including random errors of $\sigma_d/d=0.8$ and $\sigma_{\mu_b}=1$\,mas\,yr$^{-1}$, (ii) UCAC4 proper motions and RAVE DR5 distances, (iii) UCAC5 proper
motions and McMillan distances cut to
$\sigma_d/d<25\%$, (iv) the RAVE-TGAS-McMillan sample cut to
$\sigma_d/d<25\%$, and (v) the RAVE-TGAS-ABJ sample.

The leftmost column shows the pattern obtained from the mock RAVE sample including random errors in distance of $\sigma_d/d=0.8$ and proper motion of $\sigma_{\mu_b}=1$\,mas\,yr$^{-1}$ to match the errors in the RAVE-TGAS-McMillan sample. The obtained pattern in the Galaxia model shows amplitudes that are similar to the data, however, the signature of the pattern does not match any of the structures in our samples. Therefore, the patterns observed with the RAVE-TGAS subsamples may in fact be due to non-axisymmetries in the Milky Way disc.

In all RAVE-TGAS subsamples (3\textsuperscript{rd}, 4\textsuperscript{th} and
5\textsuperscript{th} column), the top panels, with distances $6<R<7.5$\,kpc,
show positive vertical velocities in the northern Galactic hemisphere and
negative in the southern hemisphere. Thus, the stellar motions agree closely
with those expected for a breathing mode. This is consistent with the
velocity trend observed in Figure 4 of \citet{Widrow2012}. However, the
bottom panels covering the distance range $7.5<R<9$\,kpc are mainly dominated
by negative values of $V_z$ above and below the plane, corresponding to a
bending mode. This signal is clear in the last column, based on ABJ
distances, but weak in the other two columns. From top to bottom we see a gradual change from a breathing to a bending mode.

In the 2\textsuperscript{nd} column of Fig.\,\ref{fig:Widrowcompare} we see with weak
significance the opposite behaviour: a gradual change from a bending mode at
distances $6<R<7$\,kpc to a breathing mode at distances $7<R<9$\,kpc. Given
the weakness of this signal and the inferior quality of both the distances
and the proper motions on which the leftmost column is based, we discount
this combination of breathing and bending modes.

In contrast to previous results suggesting a breathing mode perturbation in
the extended solar neighbourhood, our analysis supports a combination of
breathing mode inside $R_0$ and bending mode outside $R_0$. Inwards of the
solar radius, a bar and/or spiral perturbations could have induced the
breathing mode; alternatively, this could be due to a satellite
perturbation. The expected vertical variations due to the spiral structure are of the order of $\approx 4$\,km\,s$^{-1}$ \citep{2014Faure} and just a small contribution due to the bar $0.5$\,km\,s$^{-1}$ \citep{Monari2015}, thus the vertical amplitude we observe of $\approx 10$\,km\,s$^{-1}$ suggests that the vertical structure at $R<R_0$ may not have a purely internal origin but arise from the combination of internal and external mechanisms. On the other hand, the bending mode outside the solar radius is consistent
with an external perturbation, for example caused by the Sagittarius dwarf
galaxy or a dark matter subhalo passing through the disc. \citet{Gomez2013} studied a high-resolution simulation of the interaction between the Milky Way and Sgr and showed that the amplitude of the induced mean vertical velocity is $\sim8$\,km\,s$^{-1}$. Using a similar simulated setup, \citet{Elena2016} reported vertical streaming motions of $10-20$\,km\,s$^{-1}$. Both of these are consistent with our findings. The structure we find is likely related to the Galactic warp extending outside the solar radius.

The possible combination of breathing and bending mode
could thus be seen as a superposition of waves existing simultaneously
in the Milky Way disc. However, further modelling is needed to understand the individual effects that create the breathing-bending mode perturbation. This will be the subject of a forthcoming paper.
\section{Summary and Conclusions} \label{sec:Concl} 

We have used RAVE data combined with the Tycho-Gaia astrometric solution
catalogue (TGAS) to study the evolution of median velocities with Galactic
radius and position relative to the plane. We studied two main samples: (i)
a sample obtained from RAVE DR5 cross-matched with the Tycho\,-2, PPMXL and
UCAC4 catalogues (RAVE DR5 sample); (ii) samples comprising stars in RAVE DR5
that have TGAS astrometry using distances either inferred from TGAS
trigonometric parallaxes by \citet{Astraatmadja} or distances obtained by
\citet{McMillan2017} by combining TGAS parallaxes with spectrophotometric
data. In agreement with recent studies, we identified asymmetries
in $V_R$ and $V_z$ and less pronounced asymmetries in $V_{\phi}$. Independently of the proper motion catalogue used, we observed a somewhat symmetrical decline in $V_{\phi}$ with increasing $|z|$.

At distances $\left | z \right |<1$\,kpc, we confirmed the gradient in
Galactocentric radial velocity $V_R$ previously observed. In our RAVE DR5
sample, using PPMXL proper motions, we found the largest gradient in the
southern Galactic hemisphere to be $\partial V_R /\partial R = -5.99\pm
0.81$\,km\,s$^{-1}$\,kpc$^{-1}$, consistent with the gradient reported by
\citet{siebertgradient}. In the northern hemisphere at $R<R_0$, we identified
similar gradients, with the largest value being $\partial V_R /\partial R=
-8.53\pm 1.82$\,km\,s$^{-1}$\,kpc$^{-1}$ obtained with the Tycho\,-2 proper
motions. Using our more accurate RAVE-TGAS-McMillan sample, we measured 
$\partial V_R /\partial R= -7.01\pm 0.61$\,km\,s$^{-1}$\,kpc$^{-1}$ in the
southern Galactic hemisphere and $\partial V_R /\partial R= -9.42\pm
1.77$\,km\,s$^{-1}$\,kpc$^{-1}$ in the northern hemisphere. As seen in Fig.\,\ref{fig:RAVE_TGAS_ABJ}, middle panels, the gradient seems to reverse sign at $R>R_0$, consistent with the expected effect of spiral structure (\citealp{2014Faure}; \citealp{2016Monari}; this, however, does not mean that the bar plays no role in shaping the gradient). In both the RAVE DR5 and RAVE-TGAS samples, we used the projection of the
line-of-sight velocity on the Galactic plane in the direction of the Galactic
centre and anticentre to compare our results with the radial gradient
previously obtained. Our samples proved to be consistent within $1\sigma$. While our sample and those of many previous works is a mixture of stars with different ages and chemistry, the radial gradient can vary strongly for stellar populations with narrow ranges in ages and/or metallicity (Wojno et al. Submitted).

The $V_z$ velocity fields display a more complex structure, the origin of
which is under debate. Spiral arms (\citealp{2014Faure};
\citealp{2016Monari}), and to a lesser extent the bar \citep{Monari2015},
naturally induce a pattern with odd parity in the $V_z$ distribution of stars with respect to $z$, associated with a breathing mode. In contrast, bending modes (even parity in
the $V_z$ distribution) are attributed mainly to external perturbations such
a satellite galaxy or dark matter subhalo crossing the Galactic
plane. \citet{Widrow2014}, however, showed that a passing satellite galaxy
could produce both bending and breathing modes depending on the vertical
velocity of the satellite.

We have shown that maps of $V_z$ depend strongly on the adopted 
proper motions and distances. \citet{wobbly} and
\citet{GK2013} employed a compilation of proper motions and found that inside
the solar radius, there was upward motion above the plane and
downward motion below it. Outside the solar radius this velocity pattern
reversed. This pattern is consistent with that of a breathing mode.

Using our RAVE DR5 sample, with more accurate proper motions from the UCAC4
catalogue, we confirmed the observed breathing mode outside the solar radius.
However, inside $R_0$, we found upward motion both above and
below the plane, consistent with a bending mode. Thus using the UCAC4 proper
motions with RAVE DR5 changes the breathing mode observed by Williams/Kordopatis
to a possible combination of bending and breathing modes (compare right column
panels of Fig.\,\ref{fig:DR5_vz}).

Our most accurate velocity field, obtained with TGAS proper motions and
McMillan distances, which exploits both TGAS parallaxes and the
Infrared Flux Method, supports a combination of bending and breathing modes.

After studying the components of the vertical velocity, we found the Williams/Kordopatis
breathing mode to be due to the contribution of the transverse velocity component dominating
that of line-of-sight velocities (Fig.\,\ref{fig:GK_comp}). The structure resulting from the transverse velocity alone depends mainly on the proper motion, while the distance increases its amplitude. By adding
artificial distance errors to our RAVE-TGAS sample and to an axisymmetric mock RAVE-like
sample from Galaxia, we were able to increase the transverse velocity in both
samples and artificially reproduce the breathing mode observed by \citetalias{wobbly}
(see Fig.\,\ref{fig:TGAS_Galaxia}).

The combination of modes found in the RAVE-TGAS sample is the inverse of that
found with UCAC4 proper motions.
 Inside $R_0$ we identified patterns that appear consistent with a breathing mode
and outside $R_0$ they seem to agree with a bending mode. Similar structure is found with
UCAC5 proper motions (see Fig.\,\ref{fig:Widrowcompare}). This combination of
breathing and bending modes could be seen as a superposition of waves
existing simultaneously in the Milky Way disc. Unlike the pattern found with
UCAC4, the TGAS/UCAC5 pattern makes physical sense, inwards of the solar
radius, a bar and/or spiral perturbations could induce the observed breathing
mode, while outside, bending modes would likely be generated by external
perturbations such as a passing satellite galaxy or a dark matter subhalo.
Further modelling is needed to understand what combination of perturbations
to the Milky Way disc can induce a bending/breathing mode outside/inside the
solar circle.

To confirm our results that the Milky Way exhibits both bending and breathing modes, a larger disc volume must be probed. Compared to {\it Gaia} DR1, {\it Gaia} DR2 will cover a significantly larger volume of the Milky Way disc and improve significantly the data systematics. For stars within the RAVE magnitude, {\it Gaia} DR2 will have a preliminary parallax error of $\sigma_{\varpi}=0.03$\,mas \citep{Katz2017} and proper motion uncertainties of $\sigma_{\mu}\approx 0.04$\,mas\,yr$^{-1}$ \citep{Marchetti2017}. This will improve the median $\sigma_{V_z}$ and $\sigma_{V_r}$ by a factor of $6$. Thus, using data from {\it Gaia} DR2, may solve the question whether the Milky Way is still just breathing.

\section*{Acknowledgements}
I. Carrillo is grateful to Ralf-Dieter Scholz for the many discussions about proper motions, to Mary E. K. Williams for comments that improved the quality and clarity of this paper, to  Lawrence M. Widrow and Matthew H. Chequers for useful discussions about breathing and bending  modes, to Jorrit Hagen for discussions on the effects of distance errors and to Haifeng Wang for discussions on velocity asymmetries. Funding for RAVE has been provided by: the Australian Astronomical Observatory; the Leibniz-Institut fuer Astrophysik Potsdam (AIP); the Australian National University; the Australian Research Council; the French National Research Agency; the German Research Foundation (SPP 1177 and SFB 881); the European Research Council (ERC-StG 240271 Galactica); the Istituto Nazionale di Astrofisica at Padova; The Johns Hopkins University; the National Science Foundation of the USA (AST-0908326); the W. M. Keck foundation; the Macquarie University; the Netherlands Research School for Astronomy; the Natural Sciences and Engineering Research Council of Canada; the Slovenian Research Agency (research core funding No. P1-0188); the Swiss National Science Foundation; the Science \& Technology Facilities Council of the UK; Opticon; Strasbourg Observatory; and the Universities of Groningen, Heidelberg and Sydney.
The RAVE web site is at https://www.rave-survey.org. E.K.G.\ acknowledges support by Sonderforschungsbereich ``The Milky Way System'' (SFB 881) of the German Research Foundation (DFG), particularly through subproject A5.

This work has made use of data from the European Space Agency (ESA)
mission {\it Gaia} (\url{http://www.cosmos.esa.int/gaia}), processed by
the {\it Gaia} Data Processing and Analysis Consortium (DPAC,
\url{http://www.cosmos.esa.int/web/gaia/dpac/consortium}). Funding
for the DPAC has been provided by national institutions, in particular
the institutions participating in the {\it Gaia} Multilateral Agreement.
This work was supported by
the European Research Council under the European Union's Seventh Framework
Programme (FP7/2007-2013)/ERC grant agreement no.~321067.




\bibliographystyle{mnras}
\bibliography{mine2} 




\appendix




\bsp	
\label{lastpage}
\end{document}